\documentclass[a4paper,11pt]{article}
\pdfoutput=1 \usepackage{amssymb}
\usepackage{amsmath}
\usepackage{braket}
\usepackage{mathtools}
\usepackage[usenames,dvipsnames,svgnames,table]{xcolor}
\usepackage [utf8] {inputenc}
\usepackage{color}
\usepackage{cite}
\usepackage{subcaption}
\usepackage{lmodern}
\usepackage{footnote}
\usepackage{glossaries-extra}
\setabbreviationstyle[acronym]{long-short}
\glssetcategoryattribute{acronym}{nohyperfirst}{true}

\usepackage{slashed}
\usepackage[pdftex] {graphicx}
\usepackage{multirow}
\usepackage{jheppub} 
\usepackage{here}
\usepackage{hyperref}
\usepackage{verbatim}
\usepackage[justification=centering, singlelinecheck=false]{caption}
\usepackage{cleveref}
\usepackage{listings}
\usepackage{fancyvrb}

\Crefname{equation}{Eq.}{Eqs.}
\Crefname{section}{Sec.}{Secs.}

\newcommand{\cD}[0]{\mathcal D}

\newcommand{\cI}[0]{\mathcal I}
\newcommand{\cJ}[0]{\mathcal J}
\newcommand{\cK}[0]{\mathcal K}

\newcommand{\cM}[0]{\mathcal M}
\newcommand{\cO}[0]{\mathcal O}

\newcommand{\cS}[0]{\mathcal S}

\newcommand{\cY}[0]{\mathcal Y}

\newcommand{\wt}[0]{\widetilde}
\newcommand{\wh}[0]{\widehat}

\newcommand{\df}[0]{\mathrm{df}}
\newcommand{\iso}[0]{{\rm iso}}

\newcommand{\Kdf}[0]{{\cK_{\df,3}}}

\newcommand{\PV}[0]{{\mathrm{PV}}}
\newcommand{\Mdf}[0]{\mathcal{M}_{\df,3}}
\newcommand{\K}[0]{\mathcal K}

\newcommand{\chpt}{$\chi$PT}
\newcommand{\p}{\partial }

\newcommand{\bm}[0]{\boldsymbol}

\newcommand{\KB}[0]{{\cK^{B,1}_{\df,3}}}
\newcommand{\KE}[0]{{\cK^{E,1}_{\df,3}}}

\newcommand{\KSS}[0]{Kim:2005gf}

\newcommand{\HSQCa}[0]{Hansen:2014eka}
\newcommand{\HSQCb}[0]{Hansen:2015zga}

\newcommand{\HSTH}[0]{Hansen:2016fzj}

\newcommand{\BHSQC}[0]{Briceno:2017tce}
\newcommand{\BHSnum}[0]{Briceno:2018mlh}
\newcommand{\BHSK}[0]{Briceno:2018aml}
\newcommand{\dwave}[0]{Blanton:2019igq}

\newcommand{\largera}[0]{Romero-Lopez:2019qrt}

\newcommand{\RtoK}[0]{Jackura:2019bmu}
\newcommand{\HHanal}[0]{Blanton:2019vdk}
\newcommand{\isospin}[0]{Hansen:2020zhy}

\newcommand{\threehadrons}[0]{Blanton:2021llb}
\newcommand{\BSQC}[0]{Blanton:2020jnm}
\newcommand{\BSequiv}[0]{Blanton:2020gha}
\newcommand{\BSnondegen}[0]{Blanton:2020gmf}
\newcommand{\BStwoplusone}[0]{Blanton:2021mih}

\newcommand{\Akakia}[0]{Hammer:2017uqm}
\newcommand{\Akakib}[0]{Hammer:2017kms}
\newcommand{\AkakiTHa}[0]{Pang:2019dfe}
\newcommand{\AkakiTHb}[0]{Muller:2020vtt}

\newcommand{\DDK}[0]{Pang:2020pkl}
\newcommand{\AkakiRel}[0]{Muller:2021uur}
\newcommand{\MDpi}[0]{Mai:2018djl}
\newcommand{\MD}[0]{Mai:2017bge}
\newcommand{\MDHH}[0]{Mai:2019fba}
\newcommand{\Akakinum}[0]{Doring:2018xxx}
\newcommand{\Maiaone}[0]{Mai:2021nul}

\newcommand{\ThreeQCDNumerics}[0]{%
Mai:2018djl,
Horz:2019rrn, 
Blanton:2019vdk, 
Mai:2019fba, 
Culver:2019vvu, 
Fischer:2020jzp, 
Hansen:2020otl, 
NPLQCD:2020ozd,
Alexandru:2020xqf, 
Brett:2021wyd, 
Blanton:2021llb, 
Mai:2021nul} 

\newcommand{\ThreeBody}[0]{%
Briceno:2012rv,
Polejaeva:2012ut,
Hansen:2014eka,
Hansen:2015zga,
Briceno:2017tce,
Hammer:2017uqm,
Hammer:2017kms,
Mai:2017bge,
Briceno:2018mlh,
Briceno:2018aml,
Blanton:2019igq,
Pang:2019dfe,
Jackura:2019bmu,
Briceno:2019muc,
Romero-Lopez:2019qrt,
Hansen:2020zhy,
Blanton:2020gha,
Blanton:2020jnm,
Pang:2020pkl,
Hansen:2020otl,
Romero-Lopez:2020rdq,
Blanton:2020gmf,
Muller:2020vtt,
Blanton:2021mih,
Muller:2021uur}

\newcommand{\IntegralEquations}[0]{%
Jackura:2020bsk,
Hansen:2020otl,
Mai:2021nul,
Dawid:2021fxd}

\newcommand{\ThreeBodyReviews}[0]{%
Hansen:2019nir,
Rusetsky:2019gyk,
Mai:2021lwb,
Romero-Lopez:2021zdo}

\newacronym{CMF}{CMF}{center-of-momentum frame}

\DeclareFixedFont{\ttb}{T1}{txtt}{bx}{n}{9} 
\DeclareFixedFont{\ttm}{T1}{txtt}{m}{n}{9}  

\definecolor{deepblue}{rgb}{0,0,0.5}
\definecolor{deepred}{rgb}{0.6,0,0}
\definecolor{deepgreen}{rgb}{0,0.5,0}

\newcommand\pythonstyle{\lstset{
language=Python,
basicstyle=\ttm,
morekeywords={self},              
keywordstyle=\ttb\color{deepblue},
emph={MyClass,__init__},          
emphstyle=\ttb\color{deepred},    
stringstyle=\color{deepgreen},
frame=tb,                         
showstringspaces=false
}}

\lstnewenvironment{python}[1][]
{
\pythonstyle
\lstset{#1}
}
{}

\newcommand\pythoninline[1]{{\pythonstyle\lstinline!#1!}}

\preprint{}

\title{Implementing the three-particle quantization condition for $\pi^+\pi^+K^+$ and related systems }
\author[1]{Tyler D. Blanton}
\affiliation[1]{Department of Physics, University of Maryland, College Park, MD 20742, USA}

\author[2,3]{, Fernando Romero-L\'opez}
\affiliation[2]{CTP, Massachusetts Institute of Technology, Cambridge, MA 02139, USA}
\affiliation[3]{IFIC, CSIC-Universitat de Val\`encia, 46980 Paterna, Spain}

\author[4]{, and  Stephen R. Sharpe}
\affiliation[4]{Physics Department, University of Washington, Seattle, WA 98195-1560, USA}

\emailAdd{blanton1@umd.edu}
\emailAdd{fernando@mit.edu}
\emailAdd{srsharpe@uw.edu}

\abstract{
 Recently, the formalism needed to relate the finite-volume spectrum of systems
of nondegenerate spinless particles has been derived.
In this work we discuss a range of issues that arise when implementing this formalism
in practice, provide further theoretical results that can be used to check the implementation,
and make available codes for implementing the three-particle quantization condition.
Specifically, we discuss the need to modify the upper limit of the cutoff function due to the fact that
the left-hand cut in the scattering amplitudes for two nondegenerate particles moves closer to threshold;
we describe the decomposition of the three-particle amplitude $\Kdf$ into the matrix basis
used in the quantization condition, including both $s$ and $p$ waves, with the latter arising
in the amplitude for two nondegenerate particles;
we derive the threshold expansion for the lightest three-particle state in the rest frame up to
$\cO(1/L^5)$;
and we calculate the leading-order predictions in chiral perturbation theory for $\Kdf$ in the
$\pi^+\pi^+K^+$ and $\pi^+K^+K^+$ systems.
We focus mainly on systems with two identical particles plus a third that is different 
(``2+1'' systems).
We describe the formalism in full detail, and present numerical explorations in toy models,
in particular checking that the results agree with the threshold expansion, and making a prediction
for the spectrum of $\pi^+\pi^+K^+$ levels using the two- and three-particle interactions predicted
by chiral perturbation theory.
}

\allowdisplaybreaks

\preprint{MIT-CTP/5360}
\begin{document}

\maketitle
\flushbottom
\clearpage

\section{Introduction}
 \label{sec:intro}
%

In the last few years, considerable progress has been made in the development of the formalism
needed to connect the three-particle finite-volume spectrum in quantum field theories
to infinite-volume scattering amplitudes~\cite{\ThreeBody}, 
and in the implementation of this formalism and its application to the results from
lattice QCD (LQCD) simulations~\cite{\ThreeQCDNumerics}.
For recent reviews see Refs.~\cite{\ThreeBodyReviews}.
In particular, the recent application of the formalism to the $3\pi^+$ and $3 K^+$ systems
in Ref.~\cite{\threehadrons}
demonstrates that it is possible to extract three-particle contact interactions
given enough precisely determined spectral levels.

In this paper we discuss the implementation of the recent generalizations of the formalism
to particles that are not identical.
Most of our focus is on ``2+1'' systems, 
i.e.~those consisting of two identical particles together with a 
third, different, particle~\cite{\BStwoplusone},
but we also consider completely nondegenerate three-particle systems,
for which the formalism was derived in Ref.~\cite{\BSnondegen}.
The simplest examples of nondegenerate systems in QCD---namely 
$\pi^+\pi^+ K^+$ and $\pi^+ K^+ K^+$---represent the next step in terms of complication
after the $3\pi^+$ and $3K^+$ systems. The interactions are still repulsive, and there are
no three-particle resonances, 
but there is a pair of two-particle subchannels,
e.g.~$2\pi^+$ and $\pi^+ K^+$ when considering $\pi^+\pi^+ K^+$ .
Furthermore, the $\pi^+K^+$ channel has interactions in both even and odd partial waves,
unlike for identical pairs for which only even partial waves contribute.
The resulting formalism is more complicated than that for three identical particles, having
an additional flavor index corresponding to the two subchannels.

The presentation in Refs.~\cite{\BSnondegen,\BStwoplusone} 
was focused on the derivation of the formalism, 
with many details concerning the implementation not considered or discussed. 
Since results for the $\pi^+\pi^+ K^+$ and $\pi^+ K^+ K^+$ spectra from LQCD will be available very
soon, we think it important to pull together in one publication all the relevant details needed to
implement the formalism.

Another aim of this work is to provide ancillary theoretical results.
In particular, we have determined the first three nontrivial terms in the
$1/L$ expansion of energy of the threshold  three-particle state (the ``threshold expansion'')
for nondegenerate particles. This extends previous results for identical and
$2+1$ systems, and provides useful checks on our implementation of the formalism.
In addition, we have calculated the
leading order (LO) prediction in chiral perturbation theory (\chpt)
for the three-particle K matrix, $\Kdf$, 
for $\pi^+\pi^+ K^+$ and $\pi^+ K^+ K^+$  scattering.
This provides a baseline expectation for the results obtained when fitting to the finite-volume
spectra obtained using LQCD simulations.

A final motivation for this work is to provide to the community a well-tested and well-documented
python code that implements the quantization condition.
We have done so for $2+1$ systems, including both $s$- and $p$-wave two-particle interactions,
and also for identical and completely nondegenerate systems, including only $s$-wave interactions.%
\footnote{ The code used in Ref.~\cite{\threehadrons} for identical three-particle systems with both $s$- and $d$-wave
two-particle interactions has not yet been fully integrated into the repository Ref.~\cite{coderepo}, but it can be provided upon request.}
This code, deposited in Ref.~\cite{coderepo}, could serve to make
the application of the three-particle formalism more widespread.

The three-particle formalism has been derived following three different approaches:\footnote{%
See also Refs.~\cite{Guo:2017ism,Klos:2018sen,Guo:2018ibd}.
} 
(i) generic relativistic effective field theory (RFT)~\cite{\HSQCa,\HSQCb,\BHSQC,\BHSnum,\BHSK,\dwave,\largera,\isospin,\BSQC,\BSnondegen,\BStwoplusone}, 
(ii) nonrelativistic effective field theory (NREFT)~\cite{\Akakia,\Akakib,\Akakinum,\AkakiTHa,\AkakiTHb,\DDK},
 and (iii) (relativistic) finite volume unitarity (FVU) \cite{\MD,\MDpi,\MDHH,\Maiaone}.
 The equivalence of the RFT and FVU approaches, aside from some technical issues, 
 has been shown in Ref.~\cite{\BSequiv}  (see also Ref.~\cite{\RtoK}).
 We also note that a Lorentz-invariant extension 
 of the NREFT formalism has recently been obtained~\cite{\AkakiRel},
 and this is expected to be equivalent to the other approaches~\cite{Akakiprivate}.
 In this work we follow the RFT approach.

All three approaches connect the finite-volume spectrum to scattering amplitudes in two steps.
The first involves a quantization condition, an equation whose solutions give the spectrum
in terms of two- and three-particle contact interactions or K matrices. 
These latter quantities are defined in infinite volume, but
are not, in general, physical, since they depend
on the details of cutoff functions and other technical choices.
In the second step, the scattering amplitude $\cM_3$ is obtained by solving
(infinite-volume) integral equations that involve the intermediate interactions or K matrices.
These integral equations lead to an $\cM_3$ that satisfies $s$-channel unitarity, 
and correctly includes initial- and final-state interactions.
In this work we consider only the first step, i.e.~the implementation of the quantization condition.
For recent progress in solving the integral equations, see Refs.~\cite{\IntegralEquations}.

The remainder of this paper is organized as follows.
We begin, in \Cref{sec:implement}, by describing the implementation of nondegenerate quantization
conditions, focusing mainly on 2+1 systems. 
We then, in \Cref{sec:chpt}, present the calculation of $\Kdf$ for 
$\pi^+\pi^+ K^+$ and $\pi^+ K^+ K^+$  scattering.
Section~\ref{sec:numerical} describes two numerical applications of our implementations:
first, in \Cref{sec:thrnum}, a comparison with the threshold expansion for a fully nondegenerate system, 
and then, in \Cref{sec:numericaltoy}, an illustration of the impact of two- and three-particle
interactions on the $\pi^+\pi^+ K^+$ and $\pi^+ K^+ K^+$ spectra for parameters likely to
be simulated in the near term.
We summarize and conclude in \Cref{sec:conc}.

We also include four appendices. Appendix~\ref{app:details} collects some technical details
related to the implementation of the $2+1$ quantization condition;
\Cref{app:threshold} outlines the derivation of the threshold expansion for three nondegenerate
particles; and \Cref{app:KtoM} derives the relationship between $\Kdf$ and $\cM_3$
for a $2+1$ system, which is needed for the \chpt\ calculation of Sec.~\ref{sec:chpt}.
Finally, in \Cref{app:code}, we provide examples of the use of our codes.

\section{Implementing the nondegenerate three-particle formalism}
\label{sec:implement}


In this section we describe the issues that arise when one implements the 
quantization conditions for nondegenerate particles.
For concreteness, and because it is likely to be useful in the near term,
we focus on the $2+1$ formalism of Ref.~\cite{\BStwoplusone}. 

We begin with a summary of the formalism, and then
discuss (i) the need to change the cutoff functions compared to the degenerate case
(or, more precisely, the need to change the maximum of the momentum of each of the particles);
(ii) how one decomposes the three-particle K matrix
into the matrix variables appearing in the quantization condition,
including both $s$- and $p$-wave two-particle interactions; and
(iii) how to project the quantization condition onto irreducible representations (irreps)
of the appropriate finite-volume symmetry group.
We relegate some technical details to \Cref{app:details}.

Throughout this section we denote the flavor of the two identical particles as $1$, and their mass
as $m_1$, while the flavor of the solitary particle is $2$ and its mass $m_2$.
The total energy of the three-particle system at rest, assuming no interactions, would then be
\begin{equation}
M = 2 m_1 + m_2\,.
\end{equation}
We assume that the finite volume is a cubic box of length $L$, and that the fields satisfy periodic
boundary conditions. Thus finite-volume momenta are drawn from the finite-volume set, i.e.,
$\bm k = (2\pi/L) \bm n_k$ with $\bm n_k$ a vector of integers.
We are interested in determining the allowed values of the total energy $E$ for a $2+1$
system with given total spatial momentum $\bm P$ (itself a member of the finite-volume set),
and box size $L$. A useful variable in the following will be the center-of-momentum frame (CMF)
energy, $E^*=\sqrt{E^2-\bm P^2}$.

The quantization condition applies (and is derived) in the continuum limit, so no lattice
spacing, $a$, is present. This means that, strictly speaking, to apply
the formalism to the results of lattice QCD simulations, one must send $a\to0$.

\subsection{Summary of formalism and definitions for $2+1$ systems}
\label{sec:formalism}

Here we recapitulate the formalism for $2+1$ systems derived in Ref.~\cite{\BStwoplusone}.
As noted in the introduction, we consider here only the implementation of
the quantization condition, i.e., the formula relating the finite-volume spectrum to $\Kdf$.
Furthermore,  we consider only the so-called symmetric form of the quantization condition,
i.e., that in which $\Kdf$ has all the symmetries of $\cM_3$.
This is the simplest to implement, as the symmetric form of $\Kdf$ involves the smallest number of parameters.

The quantization condition is\footnote{
This is valid up to exponentially suppressed corrections, i.e., those that scale with $L$
as $\exp(-m_i L)$. We will assume throughout that such corrections can be neglected.}
\begin{equation}
\det\left[\wh F_3^{-1}(E,\bm P, L) +  \wh \cK_{\df,3}(E^*)\right] = 0\,,
\label{eq:QC3}
\end{equation}
i.e., there are finite-volume levels at the energies $E$ for which the determinant vanishes,
for the given values of the box size $L$ and total momentum $\bm P$. 
The K matrix $\wh\cK_{\df,3}$ is an infinite-volume Lorentz-invariant quantity
that does not depend on $E$, $\bm P$, and $L$ separately,
but only on the CMF energy $E^*$. We discuss it separately in \Cref{sec:Kdf}.
$\wh F_3$ is an intrinsically finite-volume object that will be defined below.
Both quantities are matrices with multiple indices, over which the determinant runs.
The indices are $k\ell m i$, and we explain these in turn.
The first three are a shorthand for $\bm k \ell m$, and these are the standard indices in
all approaches to the three-particle quantization condition~\cite{\HSQCa,\Akakib,\MD}.
They represent the variables of an on-shell, finite-volume three-particle amplitude.
One of the particles is chosen as the ``spectator,'' with momentum $\bm k$ drawn from the
finite-volume set, while the remaining pair are boosted to their CMF, where the amplitude
is decomposed into spherical harmonics, leading to the $\ell m$ indices.
Further details of this decomposition will be given in \Cref{sec:Kdf}.
The final index $i$ runs over the two choices of flavor of the spectator particle, $i=1$ and $i=2$.
We follow Ref.~\cite{\BStwoplusone} and place
carets (``hats'') on quantities to indicate that they are matrices in flavor space as well as the
usual $k\ell m$ space.


The matrix $\wh F_3$ is given by
\begin{equation}
\wh F_3 = \frac{\wh F}3 - \wh F \frac1{\wh{\overline{\cK}}{}_{2,L}^{-1} + \wh F + \wh G} \wh F\,,
\label{eq:F3hat}
\end{equation}
and is composed of the kinematical matrices $\wh F$ and $\wh G$, and the matrix
$\wh{\overline{\cK}}_{2,L}$ that contains the two-particle K matrices.
The flavor-index structure of these matrices is
\begin{align}
\wh F &= {\rm diag}\left( \wt F^{(1)}, \wt F^{(2)}\right)\,,
\label{eq:Fhat}
\\
\wh G &= 
\begin{pmatrix}\wt G^{(11)} & \sqrt2 P_L \wt G^{(12)} \\
\sqrt2 \wt G^{(21)} P_L & 0 
\end{pmatrix}\,,
\label{eq:Ghat}
\\
\wh{\overline \cK}_{2,L} &= {\rm diag} \left(
{\overline \cK}_{2,L}^{(1)}, \tfrac12 {\overline \cK}_{2,L}^{(2)}\right)\,.
\label{eq:K2Lhat}
\end{align}
Here $\wt F^{(i)}$, $\wt G^{(ij)}$, and $\overline \cK_{2,L}$ are matrices with only
$k\ell m$ indices, where the superscript $i,j$ indicates the flavor of the spectator(s).
The matrix $P_L$ is a parity factor and is given by
\begin{equation}
\left[P_L\right]_{p'\ell'm';p\ell m} = \delta_{\bm p' \bm p} \delta_{\ell' \ell} \delta_{m' m} 
(-1)^\ell \,,
\label{eq:PL}
\end{equation}
and thus multiplies odd partial waves by $-1$.
The first kinematic matrix, commonly referred to as a L\"uscher zeta function, is given by
\begin{multline}
\left[\wt F^{(i)}\right]_{p' \ell' m';p \ell m} =
\delta_{\bm p' \bm p} \frac{H^{(i)}(\bm p)}{2\omega_{p}^{(i)} L^3}
\left[ \frac1{L^3} \sum_{\bm a}^{\rm UV} - \PV \int^{\rm UV} \frac{d^3 a}{(2\pi)^3} \right] 
\\
\times \left[
\frac{\cY_{\ell' m'}(\bm a^{*(i,j,p)})}{\big(q_{2,p}^{*(i)}\big)^{\ell'}}
 \frac1{4\omega_{a}^{(j)} \omega_{b}^{(k)}
  \big(E\!-\!\omega_{p}^{(i)}\!-\!\omega_{a}^{(j)}\!-\!\omega_{b}^{(k)}\big)}
\frac{\cY_{\ell m}(\bm a^{*(i,j,p)})}{\big(q_{2,p}^{*(i)}\big)^{\ell}}
\right]
\,.
\label{eq:Ft}
\end{multline}
The flavor labels are chosen as follows: if $i=1$, then $j=1$ and $k=2$,
while if $i=2$, then $j=k=1$. The sum over $\bm a$ runs over the finite-volume set,
and both sum and integral must be regularized in the ultraviolet (UV) in the same way,
although the precise choice is not important. 
We describe our choice of regulator, and give further details of the evaluation of $\wt F^{(i)}$,
in \Cref{app:details}.
On-shell energies are exemplified by
\begin{equation}
\omega_p^{(j)} = \sqrt{ \bm p^2 + m_j^2}\,,
\end{equation}
and the third momentum is $\bm b= \bm P-\bm a-\bm p$.
The $\cY_{\ell m}$ are harmonic polynomials defined with normalization such that
\begin{equation}
\cY_{\ell m}(\bm{a})
= \sqrt{4\pi} Y_{\ell m}(\widehat{a}) |\bm a|^\ell\,.
\label{eq:harmonicpoly}
\end{equation}
We use real spherical harmonics, whose form is given in \Cref{app:details}.
The quantity $q^{*(i)}_{2,p}$ is the magnitude of the relative momenta of the pair in their CMF, 
assuming all three particles are on shell.
This depends on the total momentum $\bm P$, the spectator momentum $\bm p$, 
and the flavor of the spectator, and is given by
\begin{equation}
 \big(q_{2,p}^{*(i)}\big)^2  = \frac{\lambda(\sigma_i,m_j^2, m_k^2)}{4\sigma_i}\,, \quad
\sigma_i \equiv (E-\omega^{(i)}_{p})^2 - (\bm P-\bm p)^2\,,
\label{eq:qst}
\end{equation}
where $\lambda(a,b,c) = a^2+b^2+c^2 -2 ab -2 ac - 2 bc$ is the standard triangle function.
The momentum $\bm a^{*(i,j,p)}$ is the spatial part of the four-momentum $(\omega_a^{(j)}, \bm a)$
after a boost to the CMF of the nonspectator pair, i.e., with boost velocity
$\bm \beta_p^{(i)} = - (\bm P - \bm p)/(E-\omega_p^{(i)})$.
Finally, $H^{(i)}(\bm p)$ is a cutoff, or transition, function---see discussion in 
\Cref{sec:cutoff}.

The other kinematic function is 
\begin{equation}
\left[\wt G^{(ij)}\right]_{p \ell' m';r \ell m} = 
\frac1{2\omega^{(i)}_{p} L^3}
\frac{\cY_{\ell' m'}(\bm r^{*(i,j,p)})}{\big(q_{2,p}^{*(i)}\big)^{\ell'}}
\frac{H^{(i)}(\bm p) H^{(j)}(\bm r)}{b_{ij}^2-m_k^2}
\frac{\cY_{\ell m}(\bm p^{*(j,i,r)})}{\big(q_{2,r}^{*(j)}\big)^{\ell}}
\frac1{2\omega^{(j)}_{r} L^3}\,,
\label{eq:Gt}
\end{equation}
where $\bm p^{*(j,i,r)}$ is defined analogously to $\bm r^{*(i,j,p)}$, with the roles of
$\bm r$ and $\bm p$ (and the corresponding flavors) interchanged,
while the four-vector $b_{ij}$ is 
\begin{equation}
b_{ij} = (E-\omega_{p}^{(i)}-\omega_{r}^{(j)},\bm P-\bm p-\bm r)\,,
\label{eq:bij}
\end{equation}
and, finally, $k=2$ if $i=j=1$, while $k=1$ if $\{i,j\}=\{1,2\}$ or $\{2,1\}$.

The final matrix is defined by~\cite{\HSQCa,\BHSnum,\dwave}
\begin{align}
\left[\overline{\cK}_{2,L}^{(i)}\right]_{p\ell' m'; r \ell m} &= 
\delta_{\bm p \bm r} 2\omega^{(i)}_{r} L^3 \left[\cK^{(i)}_2(\bm r) \right]_{\ell' m';\ell m}\,,
\label{eq:K2Li}
\\
\left[\cK^{(i)}_2(\bm r)^{-1}\right]_{\ell' m';\ell m}
&=
\delta_{\ell' \ell} \delta_{m' m}  
 \frac{\eta_i}{8 \pi \sqrt{\sigma_i}} 
\left\{ q_{2,r}^{*(i)} \cot \delta_\ell^{(i)}(q_{2,r}^{*(i)}) + |q_{2,r}^{*(i)}| [1-H^{(i)}(\bm r)] \right\} \,,
\label{eq:K2}
\end{align}
where $i$ is the flavor of the spectator, and the scattering occurs between the other
two particles, with $\delta_\ell^{(i)}$ the corresponding phase shift.
If $i=1$, all waves are present, and the symmetry factor is $\eta_i=1$.
If $i=2$, the scattering is between identical particles and thus occurs only in even partial waves,
and $\eta_2=1/2$.
The two-particle K matrices are standard above threshold, but have cutoff dependence
below threshold.
In order to make our definitions clear, we note that
the effective-range expansions for the $s$- and $p$-wave phase shifts are given
in terms of scattering lengths and effective ranges by
\begin{align}
q \cot \delta_0^{(i)}(q) &= - \frac1{a_0^{(i)}} + \frac{r_0^{(i)} q^2}2 + \dots
\label{eq:EREs}
\\
q^3 \cot \delta_1^{(i)}(q) &= - \frac1{a_1^{(i)}} + \dots\,.
\label{eq:EREp}
\end{align}
We stress, however,  that other parametrizations are allowed, 
for instance one that incorporates the Adler zero in the $s$-wave channel of isospin-2 $\pi\pi$ scattering~\cite{Blanton:2019vdk}.

The dependence of $\cK_2^{(i)}$ on the cutoff function $H^{(i)}$ in \Cref{eq:K2}
is an example of the freedom we have in defining this quantity. Different choices of
$H^{(i)}$ change $\cK_2^{(i)}$, $\wt F^{(i)}$, $\wt G^{(ij)}$, and $\wh\cK_{\df,3}$, such
that the energy levels are unchanged.
In Ref.~\cite{\largera} it was noted that, for identical particles, 
there is a larger class of redefinitions of $\cK_2$ that leave the solutions unchanged. 
These redefinitions are effected by changing the PV (principal value) integral prescription.
This setup is readily generalized to the $2+1$ theory, and we find that the following
change to the two-particle K matrix is allowed
\begin{align}
\left[\cK^{(i)}_2 (\bm p)^{-1} \right]_{\ell'm';\ell m} &\longrightarrow
\left[\cK^{(i)}_2 (\bm p)^{-1} \right]_{\ell'm';\ell m}
-\delta_{\ell'\ell} \delta_{m'm}
H^{(i)}(\bm p) \frac{I_\PV^{(i,\ell)}(q_{2,p}^{*(i)2})}{32\pi}\,,
\end{align}
as long as one makes a similar change in $\wt F^{(i)}$
\begin{align}
\left[\wt F^{(i)}\right]_{p'\ell'm';p\ell m} &\longrightarrow
\left[\wt F^{(i)}\right]_{p'\ell'm';p\ell m} + \delta_{p'p}\delta_{\ell'\ell} \delta_{m'm}
\frac{H^{(i)}(\bm p)}{2\omega_p^{(i)} L^3} \frac{I_\PV^{(i,\ell)}(q_{2,p}^{*(i)2})}{32\pi}\,,
\end{align}
and an appropriate redefinition of $\wh\cK_{\df,3}$.
Here $I_\PV^{(i,\ell)}(q^2)$ is an arbitrary smooth function that can be chosen
differently for each flavor $i$ and each value of $\ell$.
This freedom can be used to allow the study of subchannel resonances
and bound states, as poles in $\cK_2^{(i)}$, which would otherwise invalidate
the derivation of the quantization condition, 
can be moved outside the kinematic range of interest~\cite{\largera}.

As it stands, the quantization condition (\ref{eq:QC3}) involves infinite-dimensional matrices.
The spectator-momentum indices are bounded by the presence
of the cutoff functions $H^{(i)}$ in $\wt F^{(i)}$ and $\wt G^{(ij)}$ (which one can show
effectively restricts the matrices $\wh{\overline{\cK}}_{2,L}$ and $\wh \cK_{\df,3}$ 
as well~\cite{\HSQCa}),
but the index $\ell$ is unbounded.
Thus, in practice, one must truncate the sum over $\ell$ by hand, 
by assuming that both $\cK_{2}^{(i)}$ and $\wh \cK_{\rm df,3}$ vanish for $\ell>\ell_{\rm max}$.
This is a natural choice close to threshold, where higher values of $\ell$ are kinematically
suppressed, but for higher energies it is an approximation whose accuracy must be checked.
In our present implementation we have set $\ell_{\rm max}=1$.
Extension to higher values of $\ell_{\rm max}$ is straightforward in principle
(see, e.g., Ref.~\cite{\dwave} for the case of $\ell_{\rm max}=2$ for identical particles)
but leads to a rapid increase in unknown K-matrix parameters~\cite{\BStwoplusone}.

One feature of the truncated quantization condition is the presence of spurious
solutions at the energies of three noninteracting particles (``free solutions'').
These solutions would be shifted from the free energies, or removed altogether,
by interactions involving higher values of $\ell$, as discussed in detail
in Ref.~\cite{\dwave}. 
These spurious solutions are easy to remove in practice, either by identifying them before evaluating the quantization condition, or
by simply ignoring all solutions at free energies.

Another source of spurious solutions are the 
factors of $q_{2,p}^{*(i)\ell}$ that are present explicitly in the
denominators of $\wt F^{(i)}$ and $\wt G^{(ij)}$, and contained implicitly in the numerators of
$\cK^{(i)}_2$ and $\wh \cK_{\df,3}$.
As explained in Ref.~\cite{\dwave}, these lead to spurious solutions 
at kinematic points such that $q_{2,p}^{*(i)2}=0$ for some choices of $\bm p$ and flavor $i$,
while leaving physical solutions unaffected.
They can be consistently removed from the quantization condition
by the following changes:
\begin{equation}
\wh F \to \wh Q \wh F \wh Q\,,\ \
\wh G \to \wh Q \wh G \wh Q\,,\ \
\wh{\overline{\cK}}_{2,L} \to \wh Q^{-1} \wh{\overline{\cK}}_{2,L} \wh Q^{-1}\,, \ \
\wh \cK_{\df,3} \to \wh Q^{-1} \wh \cK_{\df,3} \wh Q^{-1}\,,
\end{equation}
where
\begin{equation}
\left[\wh Q\right]_{p\ell' m' i;k\ell m j} = \delta_{ij} \delta_{\bm p \bm k} \delta_{\ell' \ell} \delta_{m' m}
q_{2,p}^{*(i)\ell}\,.
\end{equation}
This change also has the practical advantage of avoiding the imaginary values of
$q_{2,p}^{*(i)}$ that arise when one is using odd angular momenta and working below the
two-particle threshold. All quantities are thus real. We use this approach in practice.

\subsection{Cutoff function}
\label{sec:cutoff}

The cutoff or transition function proposed 
for the nondegenerate system in Ref.~\cite{\BSnondegen} is
\begin{align}
H^{(i)}(\bm p) &= J(z_i)\,,\quad
z_i = (1+\epsilon_H) \frac{\sigma_i}{(m_j+m_k)^2}\,,
\label{eq:Hiold}
\\
J(z) &= \left\{ 
\begin{array}{ll} 0, & z \le 0 \\ 
\exp\left(-\frac1z\exp\left[-\frac1{1-z}\right]\right), & 0 < z < 1\\
1, & 1 \le z \,. \end{array} \right.
\end{align}
Note that this is closely based on the original form for identical particles given in Ref.~\cite{\HSQCa}.
Here $i$, $j$, and $k$ are flavor labels, which are all different. In the $2+1$ case of interest,
these are drawn from $\{1, 1', 2\}$.
$\sigma_i$, defined in \Cref{eq:qst}, is the squared invariant mass 
of the noninteracting pair if the
flavor of the spectator is $i$. It equals $(m_j+m_k)^2$ at threshold,
and decreases as one drops below threshold.
With these definitions, the cutoff function
$H^{(i)}$ is zero for $\sigma_i\le 0$, increases with increasing positive $\sigma_i$,
and reaches unity when $\sigma_i=(m_j + m_k)^2/(1+\epsilon_H)$.
Choosing $\epsilon_H > 0$ ensures that the cutoff function reaches unity 
below threshold. This is, strictly speaking, needed in order that all corrections to the quantization
condition are exponentially suppressed.

\begin{figure}[htbp]
\begin{center}
\includegraphics[width=0.8\textwidth]{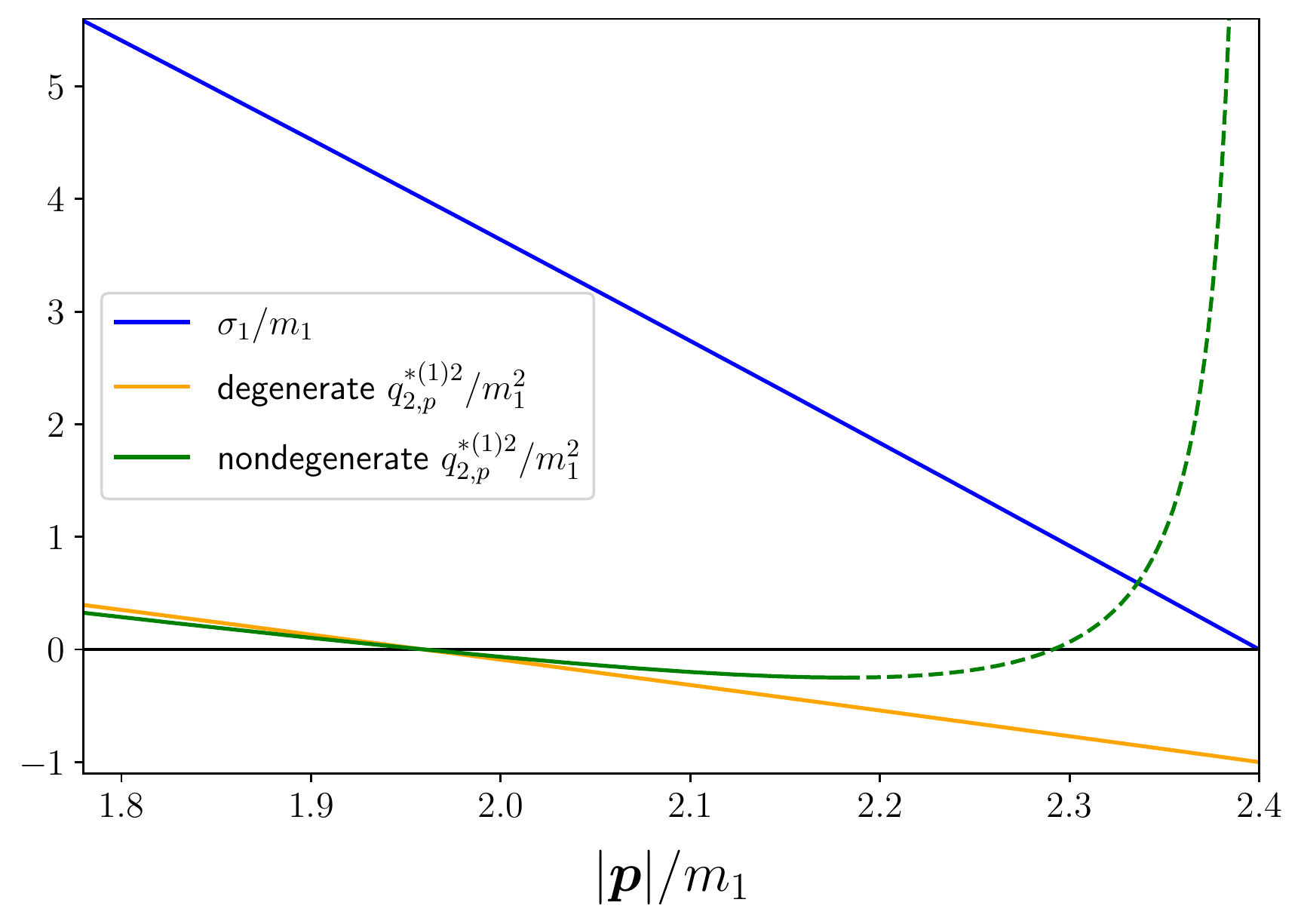}
\caption{Dependence of $\sigma_1/m_1^2$ (solid blue) and $q_{2,p}^{*(1) 2 }/m_1^2$ 
(solid orange, and green solid/dashed)
on $|\bm p|/m_1$ for $E=5 m_1$ and $\bm P=0$. Here $\bm p$ is the momentum of the
spectator, which has flavor $1$. 
The initial proposal for the cutoff function, given in \Cref{eq:Hiold}, vanishes at the
point where $\sigma_i=0$, which corresponds to the maximum value of $|\bm p|/m_1$ used.
The two curves for $q_{2,p}^{*(1) 2 }/m_1^2$ are for $m_2=m_3=m_1$  (orange curve)
and for $m_2=0.5 m_1, m_3=1.5m_1$ (green), respectively. 
The latter curve is shown as dashed after the minimum, which
occurs at the position of the left-hand cut, as discussed in the text.
\label{fig:sigmaq2}}
\end{center}
\end{figure}

This form in \Cref{eq:Hiold} turns out, however, to be unsatisfactory.
The reason for this can be seen by considering the behavior of $q_{2,p}^{*(i) 2}$ below threshold.
This quantity, given in \Cref{eq:qst}, gives the square of the relative momentum in the pair
CMF. It vanishes at threshold and initially decreases as one moves further below threshold.
Its subsequent behavior depends on whether the masses of the pair are degenerate or not.
This is illustrated in \Cref{fig:sigmaq2}: for a degenerate pair, $q_{2,p}^{*(i) 2}$ decreases
monotonically until the point where $\sigma_i=0$, which occurs at $q_{2,p}^{*(i)2}=-m_i^2$;
for a nondegenerate pair, it decreases at first, but then reaches a minimum and starts increasing,
passing through zero and eventually diverging when $\sigma_i\to 0$.
If one were to use the cutoff function in \Cref{eq:Hiold} for a nondegenerate pair,
then the full range shown in the figure would contribute.
This is problematic, however, as $q_{2,p}^{*(i)2}$ can then take on physical (positive) values,
despite being far below threshold. If one uses parametrizations such as
the effective-range expansions of \Cref{eq:EREs,eq:EREp}, 
this leads to unphysical behavior in $\cK_2$.
Indeed, we have observed that this can in turn lead to spurious bound states (poles in $\cM_2$)
far below threshold. We do stress, however, that these problems do not occur in the degenerate
case, for which the cutoff function in \Cref{eq:Hiold} is satisfactory.

The above-described problem can be avoided by lowering the cutoff in the nondegenerate case,
in such a way that $q_{2,p}^{*(i) 2}$ does not extend beyond its minimum.
This is illustrated in \Cref{fig:sigmaq2} by the transition from a solid to a dashed line for
the nondegenerate curve. Using the expression in \Cref{eq:qst}, it is straightforward
to show that the minimum occurs when $q_{2,p}^{*(i) 2}= - m_{{\rm min}, jk}^2$,
where $m_{{\rm min}, jk}$ is the smaller of $m_j$ and $m_k$, or equivalently
when $\sigma_i= |m_k^2-m_j^2|$. To implement this choice, we propose defining
\begin{align}
H^{(i)}(\bm p) &= J(z'_i)\,,
\label{eq:Hinew}
\\
z'_i &\equiv 
(1+\epsilon_H) \frac{\sigma_i - |m_j^2-m_k^2|}{(m_j+m_k)^2- |m_j^2-m_k^2|}
=
(1+\epsilon_H) \frac{\sigma_i - |m_j^2-m_k^2|}{2 m_{{\rm min}, jk}(m_j+m_k)}\,.
\label{eq:newzi}
\end{align}
which agrees with the standard form in \Cref{eq:Hiold} in the degenerate limit.
We have used the new form in practice in the numerical examples shown below.
We set $\epsilon_H=0$, since, although in theory this can lead to additional power-law volume
effects, they are suppressed in practice by the fact that the interpolation function $J(z)$
remains very close to unity for a range of $z$ below unity. For example $1-J(0.92) \approx 4 \times 10^{-6}$. 

We have described the need for the lowered cutoff in pragmatic terms. There is, however,
an additional reason for this choice that is based on the underlying physics.
This is the position of the left-hand cut in $\cK_2$, which is inherited from that in $\cM_2$.
This nonanalyticity arises because of cuts in the $t$ and $u$ channels, which occur for
subthreshold kinematics in the $s$ channel. The derivation of the quantization condition
accounts for $s$-channel cuts, but not those in $t$ and $u$ channels. Indeed, the
resultant nonanalyticities invalidate the derivation of the three-particle quantization condition,
where subthreshold two-particle contributions are important. Thus one must either deal with them
explicitly in a yet-to-be-developed manner, or place the cutoff so that they are avoided.
The $t$-channel cut occurs when $t_2= 4 m_{{\rm min},jk}^2$ 
and $s_2=u_2= |m_j^2-m_k^2|$, where the subscripts on $s_2$, $t_2$, and $u_2$ indicate
that they apply to a two-particle subchannel. The $u$-channel cut has the same position
except with $t_2 \leftrightarrow u_2$. In both cases, $\sigma_i=s_2=|m_j^2-m_k^2|$,
which we recognize as the same position as where $q_{2,p}^{*(i) 2}$ reaches its minimum.
Thus the cutoff function given in \Cref{eq:Hinew} avoids the left-hand cut, and the
quantization condition remains strictly valid.

It might be thought problematic that a lower cutoff is required for the nondegenerate theory---it
certainly conflicts with the usual notion of a UV cutoff that one can send arbitrarily large,
a point stressed recently in Ref.~\cite{\AkakiRel}.
This is why we have also called $H^{(i)}$ a ``transition function,'' because, in all
derivations in the RFT approach, it has the effect of transitioning the two-particle amplitude
that appears in the expressions between
the two-particle K matrix $\cK_2$ at threshold (where $H^{(i)}=1$)
and the two-particle amplitude $\cM_2$ far below threshold (where $H^{(i)}=0$).
As the discussion in this section has shown, the presence of the left-hand cut implies that,
within the context of a derivation that does not explicitly account for the impact of the associated
nonanalyticities, the region of the transition cannot be extended further below threshold.
We stress that there is nothing inconsistent in this situation: the fact that the cutoff lies a distance
below threshold that is set by $m_{{\rm min},jk}$ implies that the exponentially suppressed
corrections that are not controlled behave as $\exp(- m_{{\rm min},jk} L)$. This is the expected
size of such corrections, which are dropped throughout the derivation.
 In practice, when studying the $\pi \pi K$ and $\pi K K$ systems,
this implies that the cutoff, in terms of $q_{2,p}^{*(i)2}$, must be placed at the {\em same}
position as in the study of the $3\pi$ system, since the minimum mass is that of the pion in all cases.

\subsection{Implementation of threshold expansion of $\Kdf$}
\label{sec:Kdf}

In this section we describe how we determine the  form of the matrix $\wh \cK_{\df,3}$ that enters the
quantization condition [\Cref{eq:QC3}]. The starting point is the result for the infinite-volume amplitude $\Kdf$. 
We label the initial momenta as $p_1$, $p_{1'}$, and $p_2$,
and the final momenta as $p'_1$, $p'_{1'}$, and $p'_2$, 
with the subscripts indicating the flavor.
All these momenta are on shell, and the total four-momentum is $(E,\bm P)$.
Using the invariance of $\Kdf$ under Lorentz transformations,
under interchange of the two identical particles separately in the initial and final state,
and under time reversal and parity,
it was shown in Ref.~\cite{\BStwoplusone} that, to linear order in the threshold expansion,\footnote{%
The normalization of the final term differs by a factor of 2 from that in Ref.~\cite{\BStwoplusone}.}
\begin{equation}
\Kdf = \K_{\df,3}^{\iso,0} +\K_{\df,3}^{\iso,1} \Delta
+ \K_{\df,3}^{B,1} \Delta_2^{S} + \K_{\df,3}^{E,1} \wt t_{22}\,.
\label{eq:thexp}
\end{equation}
Here $\K_{\df,3}^{\iso,0}$, $\K_{\df,3}^{\iso,1}$, $\K_{\df,3}^{B,1}$, and $\K_{\df,3}^{E,1}$
are real constants, while the dimensionless kinematic variables are given by
\begin{align}
\begin{split}
\Delta &= \frac{s-M}{M^2}\,,
\quad  s = (p_1+p_{1'}+p_2)^2 = P^2\,,
\\
\Delta_2^S &= \Delta_2 +\Delta'_2 \,, 
\quad \Delta_2 = \frac{(p_1+p_{1'})^2 - 4 m_1^2}{M^2}\,,
\quad \Delta'_2 = \frac{(p'_1+p'_{1'})^2 - 4 m_1^2}{M^2}\,,
\\
\wt t_{22} &= \frac{t_{22}}{M^2} = \frac{(p_2-p'_2)^2}{M^2}\,,
\end{split}
\label{eq:kindefs}
\end{align}
where $M=2m_1+m_2$. In the threshold expansion
$\Delta\sim \Delta_2^S \sim \wt t_{22}$ are assumed small compared to unity.
As we will see explicitly below,
working to linear order in the threshold expansion implies that only $s$- and $p$-wave
contributions are present, i.e.~$\ell \le 1$ and $\ell' \le 1$.

The matrix appearing in the quantization condition is~\cite{\BStwoplusone}
\begin{equation}
\wh \cK_{\df,3} = \begin{pmatrix}
[\cK_{\df,3}]_{p \ell' m' 1;k \ell m 1} & [\cK_{\df,3}]_{p \ell' m' 1;k \ell m 2}/\sqrt2
\\
[\cK_{\df,3}]_{p \ell' m' 2;k \ell m 1}/\sqrt2 & [\cK_{\df,3}]_{p \ell' m' 2;k \ell m 2}/2
\end{pmatrix}
\,.
\label{eq:symmKdf3}
\end{equation}
Each of the four entries corresponds to a different decomposition of $\Kdf$,
differing in the flavors of the spectators.
To explain how this decomposition is defined, we consider the example of
the top-right or flavor ``12'' entry. In this case, the outgoing spectator momentum has flavor 1,
so that $\bm p=\bm p_1$, 
while the incoming spectator momentum has flavor 2, implying $\bm k=\bm p_2$.
In the final state, the remaining pair has flavors $1'$ and $2$, and the remaining
kinematic degree of freedom (for fixed $E$, $\bm P$, and $\bm p_1$) is the direction
of $\bm p'_{1'}$ when boosted to the CMF of the pair,\footnote{%
Here we are following the convention used in Ref.~\cite{\BStwoplusone} by
choosing the particle of flavor $1$ to define this direction, rather than that of
flavor $2$.}
which is denoted $\wh {a}'^*$.
Similarly, in the initial state, the pair has flavors $1$ and $1'$,
and the remaining degree of freedom is the direction
of $\bm p_1$ in the incoming pair CMF, and this is denoted $\wh a^*$.
To proceed, we first express $\Kdf$ in terms of these kinematic variables, 
and then decompose into spherical harmonics as follows
\begin{equation}
\Kdf(\bm p, \wh a'^*; \bm k, \wh a^*) = 
\sum_{\ell' m', \ell m} 4\pi Y_{\ell' m'}(\wh a'^*) [\Kdf]_{p\ell' m' 1;k \ell m 2}
Y_{\ell m}(\wh a^*)\,.
\label{eq:Kdfdecomp}
\end{equation}
This is a straightforward but tedious exercise, and we sketch its results below.
The other decompositions are obtained following the same procedure with different
choices of spectator flavors.

One immediate general result is that only even values of angular momenta can be
present if the spectator momentum has flavor $2$, because the remaining pair consists of
identical particles. Since $\ell_{\rm max}=1$ for our choice of $\Kdf$, this implies that,
if the flavor index is $2$, only $\ell=m=0$ contributions are present in the decompositions.
Only if the flavor index is $1$ can both $\ell=0$ and $1$ terms appear.

The decomposition of the first two terms in \Cref{eq:thexp} is trivial. These are isotropic,
i.e., they only depend on the total CMF energy and not on the directions of the three particles.
Thus there is no dependence on $\wh a'^*$ or $\wh a^*$ in any of the decompositions,
so only $\ell'=\ell=0$ terms appear. Given the normalization choice in \Cref{eq:Kdfdecomp},
we thus find that the nonzero contributions are
\begin{equation}
[\Kdf]_{p 0 0 1;k 0 0 1}
=
[\Kdf]_{p 0 0 1;k 0 0 2}
=
[\Kdf]_{p 0 0 2;k 0 0 1}
=
[\Kdf]_{p 0 0 2;k 0 0 2}
\supset
\K_{\df,3}^{\iso,0} +\K_{\df,3}^{\iso,1} \Delta\,.
\label{eq:Kdf3matiso}
\end{equation}
We stress that these results hold for all choices of the spectator momenta $\bm p$ and $\bm k$. 

Since the decompositions of the other two terms in $\Kdf$ require more explanation, 
we will address them in separate subsections below. 
Before doing so, however, we comment on two general issues.
The first concerns the removal of factors of $q^*$ from the quantization condition.
As noted in \Cref{sec:implement}, we can do so 
by making the replacement  $\wh \cK_{\df,3} \to \wh Q^{-1} \wh \cK_{\df,3} \wh Q^{-1}$. 
The implementation of this transformation on the decompositions described 
below is very straightforward: one simply deletes all appearances of $a^*\equiv|\bm a^*|$ and $a'^*\equiv|\bm a'^*|$.

Second, we note that a useful check of the decompositions, 
and their implementation in the code, is obtained from the
fact that the underlying quantity $\Kdf$ is Lorentz invariant. It follows that the different
blocks of $\wh \cK_{\df,3}$ in \Cref{eq:symmKdf3} are themselves invariant if we transform
$P=(E,\bm P)$, $p$, and $k$ with a common, arbitrary, boost, aside from the need to apply a Wigner
D-matrix rotation to the parts of the decomposition with $\ell'=1$ and/or $\ell=1$.
The need for these extra rotations is discussed in Sec.~VII of Ref.~\cite{\BSnondegen}.
We have checked our decompositions in this manner.

\subsubsection{Decomposition of the $\cK_{\df,3}^{B,1}$ contribution}

We begin with the flavor $22$ decomposition [the bottom-right block in \Cref{eq:symmKdf3}]
as we know that this involves only $\ell'=\ell=0$.
Since $\bm p=\bm p'_2$ and $\bm k=\bm p_2$ in this case, we have that
\begin{equation}
M^2 \Delta_2 = (P-k)^2 - 4 m_1^2 
\ \ {\rm and} \ \
M^2 \Delta'_2 =(P-p)^2 - 4 m_1^2\,.
\end{equation}
Explicitly evaluating the inner products leads to the contribution
\begin{equation}
M^2 [\cK_{\df,3}]_{p 0 0 2;k 0 0 2} \supset
2 \cK_{\df,3}^{B,1}[s -  E (\omega_p^{(2)}+\omega_k^{(2)}) 
+  \bm P\cdot (\bm p +\bm k) + m_2^2- 4 m_1^2]\,.
\label{eq:Kdf3matB22}
\end{equation}
We stress that this differs from the isotropic contribution to the same element,
given in \Cref{eq:Kdf3matiso},
because it depends not only on $s$ but also on the spectator momenta.

Next we turn to the flavor $11$ decomposition [the top-left block in \Cref{eq:symmKdf3}],
which is the most complicated. Since now $\bm p=\bm p'_1$ and $\bm k=\bm p_1$,
we have
\begin{align}
\begin{split}
M^2 \Delta_2 &= 2p_1\cdot p_{1'} - 2 m_1^2
= k\cdot p_+  + k \cdot p_- -2 m_1^2\,,
\\
M^2 \Delta'_2 &= 2p'_1\cdot p'_{1'} - 2 m_1^2
= p\cdot p'_+  + p \cdot p'_- -2 m_1^2\,,
\\
p_\pm &= p_{1'}\pm p_2\,,\quad p'_\pm = p'_{1'}\pm p'_2\,.
\end{split}
\end{align}
The $p_+$ and $p'_+$ terms are simple to evaluate given that $p_+=P-k$ and $p'_+=P-p$,
and do not depend on the pair CMF directions.
The $p_-$ and $p'_-$ terms do, however, depend on these directions, because, in their
respective pair CMFs, we have
\begin{align}
p_- &= (p_{-,0}^*,2\bm a^*)\,,\quad
p^*_{-,0} = \omega_{a^*}^{(1)}-\omega_{a^*}^{(2)}\,,
\label{eq:pmst}
\\
p'_- &= (p'^*_{-,0},2\bm a'^*)\,,\quad
p'^*_{-,0} = \omega_{a'^*}^{(1)}-\omega_{a'^*}^{(2)}\,.
\label{eq:ppmst}
\end{align}
\begin{sloppypar} \noindent
To evaluate the $k\cdot p_-$ term, for example, we must boost the lab-frame four-vector
${k=(\omega_k^{(1)},\bm k)}$ into the CMF of the initial state $1'2$ pair.
This requires a boost with velocity ${\bm \beta_k= -(\bm P-\bm k)/(E-\omega_k^{(1)})}$,
and, using the notation introduced in \Cref{sec:formalism}, leads to
\end{sloppypar}
\begin{equation}
k \xrightarrow{\textrm{boost by }\bm \beta_k} \left(\omega_{k^{*k}}^{(1)}, \bm k^{*(11'k)} \right)\,,
\qquad
\omega_{k^{*k}}^{(1)} \equiv \sqrt{m_1^2 + |\bm k^{*(11'k)}|^2}\,.
\end{equation}
Similarly, for the $p\cdot p'_-$ term, we need the boost of $p$ by $\bm \beta_p$,
\begin{equation}
p \xrightarrow{\textrm{boost by }\bm \beta_p} \left(\omega_{p^{*p}}^{(1)}, \bm p^{*(11'p)} \right)\,,
\qquad
\omega_{p^{*p}}^{(1)} \equiv \sqrt{m_1^2 + |\bm p^{*(11'p)}|^2}\,.
\end{equation}
Using these results we now have
\begin{align}
\begin{split}
M^2 \Delta_2
&= E \omega_k^{(1)} - \bm k \cdot \bm P   - 3 m_1^2  
+ \omega_{k^{*k}}^{(1)}p^*_{-,0}
- 2\bm k^{*(11'k)} \cdot \bm a^* \,,
\\
M^2 \Delta'_2 
&= E \omega_p^{(1)} - \bm p \cdot \bm P   - 3 m_1^2  
+ \omega_{p^{*p}}^{(1)} p'^*_{-,0}
- 2\bm p^{*(11'p)} \cdot \bm a'^* \,,
\end{split}
\end{align}
so that the dependence on the CMF directions is explicit.

To decompose the terms linear in $\bm a^*$ and $\bm a'^*$, we use
\begin{equation}
\bm p \cdot \bm a = \frac{4\pi}3 p a \sum_{m} Y_{1m}(\wh p) Y_{1m}(\wh a)\,.
\end{equation}
This allows us to pull out the spherical harmonics in the decomposition of the flavor
11 term that is analogous to \Cref{eq:Kdfdecomp}, and thus to read off
\begin{align}
\begin{split}
M^2 [\cK_{\df,3}]_{p 0 0 1;k 0 0 1} &\supset
\cK_{\df,3}^{B,1}[E (\omega_p^{(1)} \!+\! \omega_k^{(1)}) \!-\! (\bm p\! +\! \bm k) \cdot \bm P 
\!-\! 6 m_1^2
\!+\! \omega_{k^{*k}}^{(1)}p^*_{-,0}
\!+\! \omega_{p^{*p}}^{(1)}p'^*_{-,0}]\,,
\\
M^2 [\cK_{\df,3}]_{p 1 m' 1;k 0 0 1} &\supset
-2 \cK_{\df,3}^{B,1}  \frac{a'^*}3 \cY_{1m'}(\bm p^{*(11'p)})\,,
\\
M^2 [\cK_{\df,3}]_{p 0 0 1;k 1 m 1} &\supset
-2 \cK_{\df,3}^{B,1}  \frac{a^*}3 \cY_{1m}(\bm k^{*(11'k)})\,.
\end{split}
\end{align}
Here we are using the harmonic polynomials defined in \Cref{eq:harmonicpoly}.
We observe that there are no contributions with $\ell'=\ell=1$.

The flavor off-diagonal entries of $\wh \K_{\df,3}$ can be obtained
from the results above, and we find
\begin{align}
\begin{split}
M^2 [\cK_{\df,3}]_{p 0 0 1;k 0 0 2} &\supset
\cK_{\df,3}^{B,1}[E \omega_p^{(1)} \!-\! \bm p \cdot \bm P \!+\! s
\!-\! 2 E \omega_k^{(2)} \!+\! 2 \bm P \cdot \bm k 
\!+\! \omega_{p^{*p}}^{(1)}p'^*_{-,0} \!+\! m_2^2 \!-\! 7 m_1^2]\,,
\\
M^2 [\cK_{\df,3}]_{p 1 m' 1;k 0 0 2} &\supset
-2 \cK_{\df,3}^{B,1}  \frac{a'^*}3 \cY_{1m'}(\bm p^{*(11'p)})\,,
\\
M^2 [\cK_{\df,3}]_{p 0 0 2;k 0 0 1} &\supset
\cK_{\df,3}^{B,1}[E \omega_k^{(1)} \!-\! \bm k \cdot \bm P \!+\! s
\!-\! 2 E \omega_p^{(2)} \!+\! 2 \bm P \cdot \bm p 
\!+\! \omega_{k^{*k}}^{(1)}p^*_{-,0} \!+\! m_2^2 \!-\! 7 m_1^2]\,,
\\
M^2 [\cK_{\df,3}]_{p 0 0 2;k 1 m 1} &\supset
-2 \cK_{\df,3}^{B,1}  \frac{a^*}3 \cY_{1m}(\bm k^{*(11'k)})\,.
\end{split}
\end{align}

\subsubsection{Decomposition of the $\cK_{\df,3}^{E,1}$ contribution}

The 22 block is simple to obtain using
$t_{22}=2 m_2^2- 2 p'_2\cdot p_2$, which leads to
\begin{equation}
M^2 [\cK_{\df,3}]_{p 0 0 2;k 0 0 2} \supset
2 \cK_{\df,3}^{E,1} [m_2^2 - \omega_p^{(2)}\omega_k^{(2)} + \bm p\cdot \bm k]\,.
\end{equation}
For the flavor 11 block, we need to rewrite $t_{22}$ in terms of $p_\pm$ and $p'_\pm$,
\begin{align}
t_{33} &\to 2 m_2^2 - \frac12\left(p'_+ \cdot p_+ + p'_-\cdot p_- - p'_+ \cdot p_- - p'_- \cdot p_+\right)\,.
\end{align}
To evaluate this we need
\begin{align}
\begin{split}
p'_+ \cdot p_+ &= (P\!-\!p)\cdot (P\!-\!k) =s 
- E(\omega_p^{(1)}+\omega_k^{(1)}) + \bm P\cdot (\bm p+\bm k)
+ \omega_p^{(1)} \omega_k^{(1)} - \bm p \cdot \bm k\,,
\\
p'_+ \cdot p_- &= (P\!-\!p)_0^{*(11'k)} p^*_{-,0}
-2 (\bm P\!-\!\bm p)^{*(11'k)} \cdot \bm a^*\,,
\\
p'_- \cdot p_+ &= p'^*_{-,0}(P\!-\!k)_0^{*(11'p)}
-2 \bm a'^* \cdot (\bm P\!-\!\bm k)^{*(11'p)} \,.
\end{split}
\end{align}
Here $(P\!-\!p)_0^{*(11'k)}$ is the energy component of the four-vector $p'_+=P-p$ boosted
by $\bm \beta_k$,
while $(P\!-\!k)_0^{*(11'p)}$ is the energy component of $p_+=P-k$ boosted by $\bm \beta_p$.

To obtain $p'_-\cdot p_-$ we need to boost the four-vectors $p_-$ and $p'^*_-$,
given in \Cref{eq:pmst,eq:ppmst} in their respective CMFs,
into the lab frame. 
After doing so, we find 
\begin{equation}
p'_-\cdot p_- = \gamma_k\gamma_p(1-\bm \beta_k\cdot\bm \beta_p) p'^*_{-,0} p^*_{-,0}
+2 \bm a^* \cdot \bm V + 2 \bm a'^* \cdot \bm V' + 4 a'^*_i T_{ij} a^*_j \,,
\end{equation}
where
\begin{align}
\bm V &= \left[\bm \beta_p \gamma_p 
+ \wh \beta_k \wh \beta_k\cdot \bm \beta_p (\gamma_k\!-\!1) \gamma_p
-\bm \beta_k \gamma_k\gamma_p\right] p'^*_{-,0}\,, \nonumber
\\
\bm V' &= \left[\bm \beta_k \gamma_k 
+ \wh \beta_p \wh \beta_p\cdot \bm \beta_k (\gamma_p\!-\!1) \gamma_k
-\bm \beta_p \gamma_p\gamma_k\right] p^*_{-,0}\,,
\\
T_{ij} &=\wh\beta_{p,i} \wh \beta_{k,j} \left[\gamma_p\gamma_k \beta_p\beta_k
-\wh \beta_p\cdot \wh \beta_k (\gamma_p\!-\!1)(\gamma_k\!-\!1)\right]
- \wh \beta_{p,i} \wh \beta_{p,j} (\gamma_p\!-\!1) 
- \wh \beta_{k,i} \wh \beta_{k,j} (\gamma_k\!-\!1) 
- \delta_{ij}\,. \nonumber
\end{align}
Thus we observe the first appearance of a term linear in both $\bm a'$ and $\bm a'^*$,
which gives a contribution with $\ell'=\ell=1$.

To proceed, we note that
\begin{equation}
 a'^*_i T_{ij} a^*_j=  \frac13 \cY_{1m'}(\bm a'^*) T_{m' m} \cY_{1m}(\bm a^*)\,.
\end{equation}
where $T_{m' m}$ is defined by converting $T_{ij}$ into the spherical basis, 
i.e.~with $m=1$, $0$, and $-1$ corresponding to $j=x$, $z$, and $y$, respectively. 
Using this, and the results above, we find
\begin{align}
\begin{split}
\begin{split}
M^2 [\cK_{\df,3}]_{p 0 0 1;k 0 0 1} &\supset
\cK_{\df,3}^{E,1}\Big\{2 m_2^2 -
\frac12 \big[ s 
- E(\omega_p^{(1)}\!+\!\omega_k^{(1)}) + \bm P\cdot (\bm p\!+\!\bm k)
+ \omega_p^{(1)} \omega_k^{(1)} - \bm p \cdot \bm k 
\\
&\quad- (P\!-\!p)_0^{*(11'k)} p^*_{-,0} - (P\!-\!k)_0^{*(11'p)} p'^*_{-,0}
+ \gamma_k\gamma_p(1\!-\!\bm \beta_k\cdot\bm \beta_p) p'^*_{-,0} p^*_{-,0}
\big]\Big\}
\,,
\end{split}
\\
M^2 [\cK_{\df,3}]_{p 1 m' 1;k 0 0 1} &\supset
- \cK_{\df,3}^{E,1}  \frac{a'^*}3 \cY_{1m'}(\bm P^{*(11'p)} - \bm k^{*(11'p)}+\bm V')\,,
\\
M^2 [\cK_{\df,3}]_{p 0 0 1;k 1 m 1} &\supset
- \cK_{\df,3}^{E,1}  \frac{a^*}3 \cY_{1m}(\bm P^{*(11'k)} - \bm p^{*(11'k)}+\bm V)\,,
\\
M^2 [\cK_{\df,3}]_{p 1 m' 1;k 1 m 1} &\supset
- \cK_{\df,3}^{E,1}  \frac{2 a'^* a^*}3 T_{m' m}\,.
\end{split}
\end{align}

Finally we come to the off-diagonal terms. For the upper-right block, we need
\begin{align}
\begin{split}
t_{22} &= 2 m_2^2 - (p'_+ - p'_-)\cdot p_2
\\
&=  2 m_2^2 -(P-p)\cdot k +  p'^*_{-,0} \omega_{k^{*p}}^{(2)} - 2 \bm a'^* \cdot \bm k^{*(11'p)}\,,
\end{split}
\end{align}
where $\omega_{k^{*p}}^{(2)}$ is the energy of an on-shell particle of flavor 2
with momentum $\bm k^{*(11'p)}$.
For the lower-left block, we need
\begin{align}
t_{22} 
&=  2 m_2^2 -p \cdot (P-k) + p^*_{-,0} \omega_{p^{*k}}^{(2)}
 - 2 \bm p^{*(11'k)} \cdot \bm a^* \,.
\end{align}
We can now read off
\begin{align}
\begin{split}
M^2 [\cK_{\df,3}]_{p 0 0 1;k 0 0 2} &\supset
\cK_{\df,3}^{E,1} [2 m_2^2 - (E-\omega_p^{(1)}) \omega_k^{(2)} + (\bm P - \bm p) \cdot \bm k
+  p'^*_{-,0} \omega_{k^{*p}}^{(2)} ] \,,
\\
M^2 [\cK_{\df,3}]_{p 1 m' 1;k 0 0 2} &\supset
-\cK_{\df,3}^{E,1}  \frac{2 a'^*}3 \cY_{1m'}(\bm k^{*(11'p)})\,.
\\
M^2 [\cK_{\df,3}]_{p 0 0 2;k 0 0 1} &\supset
\cK_{\df,3}^{E,1} [2 m_2^2 - \omega_p^{(2)}(E-\omega_k^{(1)})  +\bm p \cdot (\bm P - \bm k) 
+ p^*_{-,0} \omega_{p^{*k}}^{(2)}] \,,
\\
M^2 [\cK_{\df,3}]_{p 0 0 2;k 1 m 1} &\supset
- \cK_{\df,3}^{E,1}  \frac{2 a^*}3 \cY_{1m}(\bm p^{*(11'k)})\,.
\end{split}
\end{align}

\subsection{Irrep projections}
\label{sec:projections}

In order to compare solutions of the quantization condition to a physical finite-volume spectrum obtained from lattice QCD, one must first classify the solutions into the irreps of the appropriate symmetry group of the system, namely the little group of transformations of the cube that leave the total momentum $\bm P$ invariant:
\begin{align}
	\text{LG}(\bm P)\equiv \{R\in O_h | R\bm P = \bm P \} \,,
\end{align}
where $O_h$ is the full cubic group with 48 elements.
As in previous works (e.g.~Refs.~\cite{\dwave,\HHanal}), 
we accomplish this by projecting the matrices appearing 
in the quantization condition onto individual irrep subspaces.
Unlike in those papers, however, here we have the additional complication of nondegenerate particle flavors, which adds a layer of structure to the projection matrices.

In order to interpret some of the results that we present here and in \Cref{app:GT},
it is useful to list the little groups for each of the classes of total momenta. We 
collect these, along with the irreps, in \Cref{tab:groups}. 

\begin{table}[htp]
\begin{center}
\scalebox{0.95}{
\begin{tabular}{c|c|c}
$\bm d_{\rm ref}$ & $\text{LG}(\bm P)$ & irreps\\
\hline
$(0,0,0)$& $O_h$ 	& $A_{1g}[1],A_{2g}[1],E_g[2],T_{1g}[3],T_{2g}[3],
A_{1u}[1],A_{2u}[1],E_u[2],T_{1u}[3],T_{2u}[3]$ \\
$(0,0,n)$& $C_{4v}$	& $A_1[1],\,A_2[1],\,B_1[1],\,B_2[1],\,E[2]$\\
$(n,n,0)$& $C_{2v}$ & $A_1[1],\,A_2[1],\,B_1[1],\,B_2[1]$\\
$(n,n,n)$& $C_{3v}$ 	& $A_1[1],\,A_2[1],\,E[2]$	\\
$(n_1, n_2,0)$& $C_{2}$	 & $A_1[1],\, A_2[1]$\\
$(n_1,n_1,n_2)$& $C_{2}$& $A_1[1],\,A_2[1]$\\
$(n_1,n_2,n_3)$& $C_1$ & $A_1[1] $
\end{tabular}
}
\end{center}
\caption{
Little group $\text{LG}(\bm P)$ for each type of frame, along with its irreps, each
with its dimension listed in square brackets. Frames are denoted by 
$\bm d_{\rm def}= \bm P L/(2\pi)$, taking a canonical choice for each type of frame.
The integers $n$, $n_1$, $n_2$, and $n_3$ are nonzero, with $n_1$, $n_2$, and $n_3$ being
distinct.
}
\label{tab:groups}
\end{table}
For a 2+1 system of pseudoscalars (e.g.~$\pi\pi K$, $KK\pi$) at fixed $L$ and ${P=(E,\bm P)}$, each matrix $\wh M\in\{\wh F, \wh G, \wh{\overline\cK}_{2,L}, \wh F_3, \wh\cK_{\df,3}\}$ used in the quantization condition is invariant under a set of orthogonal transformation matrices $\{\wh U(R)\}_{R\in \text{LG}(\bm P)}$:
\begin{align}
	\wh U(R)^T \wh M \wh U(R) &= \wh M \qquad \forall R\in \text{LG}(\bm P) \,,
	\\
	[ \wh U(R) ]_{p\ell'm'i ; k\ell m j} &\equiv \delta_{ij} \delta_{\bm p, R\bm k} \delta_{\ell'\ell} \Pi(R) \cD^{(\ell)}_{m'm}(R) \,,
\end{align}
where $\cD^{(\ell)}(R)$ is a real-basis Wigner D-matrix,
and $\Pi(R)$ is the parity of transformation $R$, 
i.e.~$+1$ if $R$ is a pure rotation and $-1$ otherwise.
The latter factor appears because we consider pseudoscalar mesons; 
to describe scalars one replaces $\Pi(R)\to 1$.
The feature of these matrices that is new to this work is the added flavor structure.
Note that $\wh U(R)$ is diagonal in its flavor and partial-wave indices, 
as well as block diagonal in its spectator-momentum indices, 
with the momenta in a given block all belonging to the same finite-volume ``orbit" 
$o_{\bm k}\equiv \{R\bm k | R\in \text{LG}(\bm P)\}$.

The transformation matrices furnish a (reducible) representation of $\text{LG}(\bm P)$:
\begin{align}
	\wh U(R_1R_2) = \wh U(R_1) \wh U(R_2) \quad \forall R_1,R_2\in \text{LG}(\bm P) \,, \qquad \wh U(\bm 1_3) = \bm 1 \,,
\end{align}
which can be decomposed into irreps $I$ via projection matrices
\begin{align}
	\wh P_I \equiv \frac{d_I}{[\text{LG}(\bm P)]} \sum_{R\in \text{LG}(\bm P)} \chi_I(R) \wh U(R) \,,
\end{align}
where $[\text{LG}(\bm P)]$ is the cardinality of the little group, $d_I$ is the dimension of irrep $I$, and $\chi_I(R)$ is its character.%
\footnote{The relevant character tables can be found, e.g., in Ref.~\cite{atkins1970tables}.}
Lastly, we collect the eigenvectors of $\wh P_I$ with nonzero (unit) eigenvalue into a smaller, non-square matrix $\wh P_{I,\text{sub}}$,
which projects $\wh M$ onto the lower-dimensional irrep subspace,
\begin{align}
	\wh M_{I,\text{sub}} &= (\wh P_{I,\text{sub}})^T \wh M \wh P_{I,\text{sub}} \,.
\end{align}
The eigenvalues of $\wh M_{I,\text{sub}}$ are precisely the subset of the eigenvalues of $\wh M$ 
that lie in irrep $I$.
Thus the quantization condition \Cref{eq:QC3} can be rewritten as
\begin{align}
	\det\left[\wh F_3^{-1} +  \wh \cK_{\df,3}\right]
	&= \prod_I \det\left[ \left( \wh F_3^{-1} + \wh \cK_{\df,3} \right)_{I,\text{sub}} \right] = 0 \,,
\end{align}
allowing for irrep-by-irrep comparisons between solutions to the quantization condition and the physical finite-volume spectrum.

Although the projection matrices $\{\wh P_I\}$ have the same diagonal and block-diagonal index structure as the $\{\wh U(R)\}_{R\in \text{LG}(\bm P)}$, 
the quantization condition matrix $\wh F_3^{-1} + \wh \cK_{\df,3}$ generally does not,
 mixing together different flavors, partial waves, 
 and spectator orbits in its eigenvalues and eigenvectors.
We discuss the details of eigenvalue decomposition into different irreps in \Cref{app:GT}.

We close this section by listing in \Cref{tab:littlegroups} the 
irreps (and corresponding number of eigenvalues)
affected by the nonisotropic terms in $\Kdf$ and by $\widehat{\overline{\cK}}_{2,L}$.
We also show the two-particle irreps in the two-particle quantization condition
that are affected by $\cK_2^{(i)}$, when one considers nonidentical particles and
includes both $s$ and $p$ waves.
By comparing to \Cref{tab:groups}, one can see which irreps are not affected by either of
these interactions.
We do not list the results for the cases that are trivial.
These are the two-particle quantization condition for identical particles with only $s$-wave interactions,
and the three-particle quantization condition with isotropic terms in $\Kdf$,
for both of which there is only a single nonzero eigenvalue that lies in the trivial irrep.
As shown in the table, it turns out that the nonisotropic $\cK_{\df,3}^{B,1}$ term
does not affect any of the nontrivial irreps.
These only enter in the decompositions of the $\cK_{\df,3}^{E,1}$ term (some irreps)
and of $\widehat{\overline{\cK}}_{2,L}$ (all irreps).
The fact that $\widehat{\overline{\cK}}_{2,L}$  couples to
all available irreps follows from the results that it is a diagonal matrix with,
in general, nonzero entries in all positions, and that it is proportional to the identity matrix
in each of the orbits listed in \Cref{app:GT}.
This implies that, for large enough $L$, it contains all irreps, irrespective of the maximum value
of $\ell$.

\begin{table}[htp]
\begin{center}
\begin{tabular}{c|c|c|c|c}
$\bm d_{\rm ref}$ & $\cK_2^{(1)}$  (QC2) & $\cK_{\df,3}^{B,1}$ (QC3) & $\cK_{\df,3}^{E,1}$ (QC3) &
 $\widehat{\overline{\cK}}_{2,L}$ (QC3)\\
\hline
$(0,0,0)$ 	& $A_{1g}(1),\,T_{1u}(3)$	& $A_{1u}(2)$ & $A_{1u}(2),\, T_{1g}(3)$ & all \\
$(0,0,n)$ 	& $A_1(2),\,E(2)$  	& $A_2(2)$ & $A_2(3),\,E(2)$  &all \\
$(n,n,0)$  	& $A_1(2),\,B_1(1),\,B_1(1)$	& $A_2(2)$ & $A_2(3),\,B_1(1),\,B_2(1)$ &all \\
$(n,n,n)$ 	& $A_1(2),\,E(2)$	& $A_2(2)$ & $A_2(3),\,E(2)$ &all \\
$(n_1,n_2,0)$ 	& $A_1(3),\,A_2(1)$	& $A_2(2)$ & $A_2(4),\,A_1(1)$ & all\\
$(n_1,n_1,n_2)$ & $A_1(3),\,A_2(1)$ & $A_2(2)$ & $A_2(4),\,A_1(1)$ & all\\
$(n_1,n_2,n_3)$ & $A_1(4)$ & $A_1(2)$ & $A_1(5)$ & all\\
\end{tabular}
\end{center}
\caption{
Irrep decompositions of the eigenvalues of two- and three-particles
K matrices for different frames, and for both two- and three-particle
quantization conditions (denoted QC2 and QC3, respectively).
Frames are denoted as in \Cref{tab:groups}.
Results assume both $s$- and $p$-wave dimers in the flavor-1 spectator channel 
and only $s$ waves in the flavor-2 channel,
and are for the $L\to\infty$ limit. For finite $L$ and energies in the range of validity of
the quantization condition, some of the eigenvalues may be absent; see \Cref{app:GT} for further discussion.
The notation $I(n)$ denotes that there are $n$ nonzero eigenvalues in irrep $I$,
a number that includes the multiplicity of the irrep.
In assigning irreps, all particles are assumed to be pseudoscalars, 
so that, for example, the trivial irreps of two- and three-particle systems 
are the parity complements of each other.
The description ``all" in the final column indicates that there are nonzero eigenvalues
in every irrep that is present for a given choice of $E$, particle masses, and $L$.
Which irreps are present can be determined from the tables in \Cref{app:GT}.
We observe that, aside from the final column,
the total number of nonzero eigenvalues of each K matrix is independent of frame,
 as expected of a Lorentz-invariant object.
}
\label{tab:littlegroups}
\end{table}

\section{$\Kdf$ from chiral perturbation theory}
\label{sec:chpt}

The goal of this section is to work out the leading-order (LO)  
prediction in ChPT for $\Kdf$ for the $2+1$ systems of interest for this work: 
$\pi^+\pi^+ K^+$ and $\pi^+K^+ K^+$. 
This is a generalization of the calculation for $3\pi^+$ carried out in Ref.~\cite{Blanton:2019vdk},
and involves only minor additional technical complications arising from
the presence of nonidentical particles.

In fact, $\Kdf$ cannot be directly calculated in ChPT.
Instead, one must calculate the physical three-particle scattering amplitude
$\cM_3$, and then use the relation between $\cM_3$ and $\Kdf$.
In general, the latter involves solving integral equations,
but at LO in ChPT the relation only involves subtracting certain divergent terms from $\cM_3$.
The details of this relation for identical particles are given in Ref.~\cite{\HSQCb},
and the generalization to the 2+1 system is outlined in Ref.~\cite{\BStwoplusone}.
In \Cref{app:KtoM} we present a detailed description, the result of which is that
\begin{equation}
\cK_{\df,3}^{\rm LO} = \cM_\text{df,3}^{\rm LO} \equiv \cM_3^{\rm LO} - \cD^{\rm LO} \,,
\label{eq:MtoKLO}
\end{equation}
where the subtraction term $\cD^{\rm LO}$ is given in \Cref{eq:subLO}.

We use the standard $N_f=3$ ChPT Lagrangian:
\begin{align}
&\mathcal{L} = \frac{F^2}{4} {\rm tr}\left[ \p_\mu U  \p_\mu U^\dagger  \right] + \frac{B F^2}{2 } 
{\rm tr} \left[ \hat m_q\left( U  + U^\dagger \right)  \right]\,, 
\label{eq:chptL}
\end{align} 
with $U = \exp \left(i {\phi}/{F} \right)$, $F\approx 93\;$MeV the pseudoscalar decay constant, and
\begin{align}
 \phi = \begin{pmatrix}
\pi^0+\frac{1}{\sqrt{3}} \eta & \sqrt{2} \pi^+& \sqrt{2} K^+ \\
\sqrt{2} \pi^- & -\pi^0+\frac{1}{\sqrt{3}} \eta & \sqrt{2} K^0 \\
\sqrt{2} K^- & \sqrt2 \overline K^0 & -\frac{2}{\sqrt{3}} \eta
\end{pmatrix}\,.
\label{eq:phimatrix}
\end{align} 
We work in the isosymmetric limit, in which $\hat m_q = \text{diag }(m_l, m_l, m_s)$. 
The LO pion and kaon masses are given by
\begin{equation}
M_\pi^2 = 2 m_l B\,, \qquad M_K^2= (m_l+m_s) B\,.
\end{equation}

\subsection{Calculation of $\cM_3$}

We use the same notation as in \Cref{sec:Kdf},
labeling the incoming and outgoing momenta $p_i$ and $k_i$, respectively, with $i=1,1',2$. 
The two identical particles correspond to $i=1$ and $1'$, with mass $m_1$,
while the third corresponds to $i=2$, with mass $m_2$.
 Thus $(m_1,m_2) = (M_\pi, M_K)$ and $(M_K,M_\pi)$ correspond respectively 
 to the $\pi^+\pi^+ K^+$ and $\pi^+K^+ K^+$ systems.

\begin{figure}[h!]
  \centering  
\begin{subfigure}{.328\textwidth}
  \centering
  \includegraphics[width=0.8\linewidth]{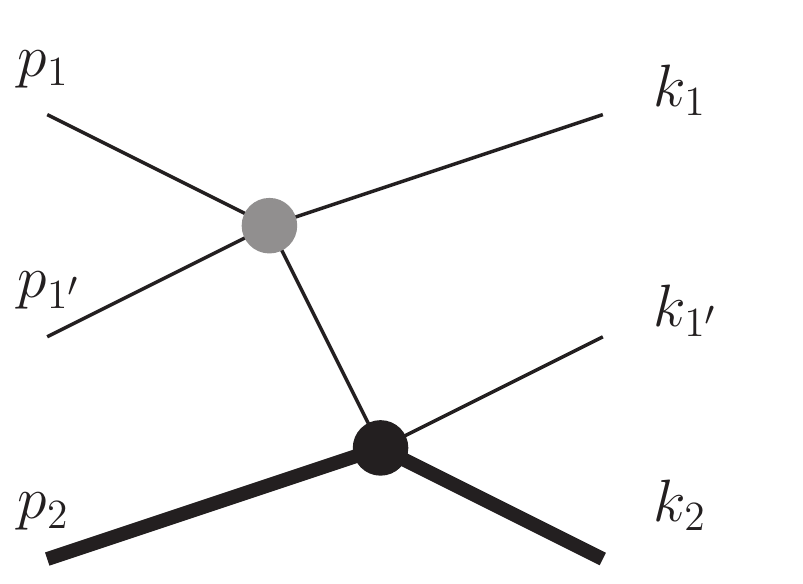}  
  \caption{}
  \label{fig:diag1}
\end{subfigure}
\begin{subfigure}{.328\textwidth}
  \centering
  \includegraphics[width=0.8\linewidth]{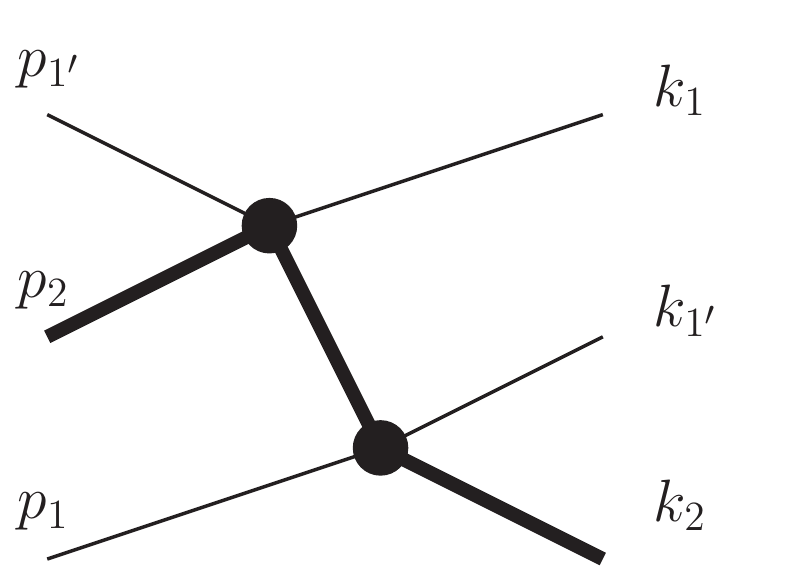}  
  \caption{}
  \label{fig:diag2}
\end{subfigure}
\begin{subfigure}{.328\textwidth}
  \centering
  \includegraphics[width=0.8\linewidth]{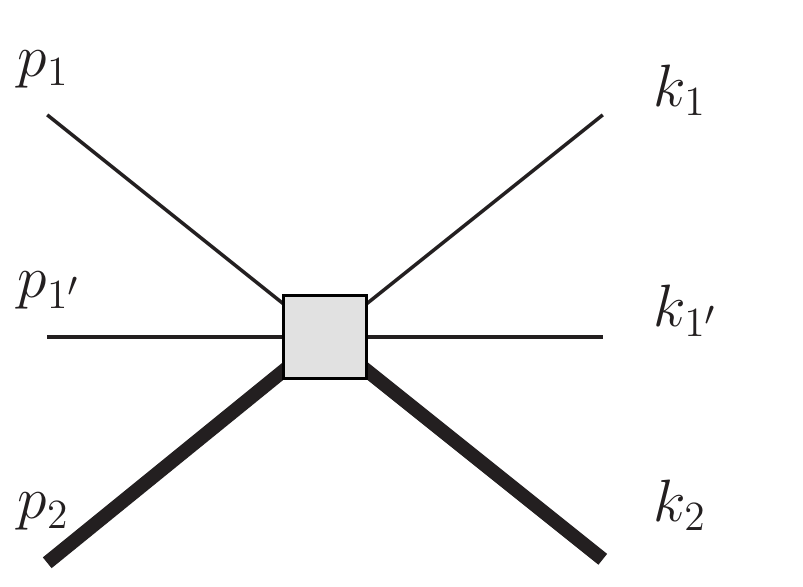}  
  \caption{}
  \label{fig:diag3}
\end{subfigure}
\caption{Diagrams contributing to $\cM_3^{\rm LO}$. Thin (thick) lines indicate particles of flavor 1 (2). 
Four-particle vertices are denoted by circles: black if the particles are of different flavors,
grey if all particles are of flavor 1. 
The gray square indicates the six-particle vertex.
}
\label{fig:diagrams}
\end{figure}

There are three different types of diagrams that contribute, as shown in \Cref{fig:diagrams}. 
The first two are one-particle exchange (OPE) diagrams, 
in which the exchanged particle can either be of flavor 1 [diagram (a)] or flavor 2 [diagram (b)]. 
Both OPE diagrams appear four times, with different momentum labels or ordering of vertices.
In addition, there is a contact term resulting from the six-meson vertex [diagram (c)].

As can be seen explicitly by expanding the chiral Lagrangian in \Cref{eq:chptL}, 
the terms contributing to $\pi^+ \pi^+$ and $K^+ K^+$ scattering are formally identical up 
to $\pi \leftrightarrow K$ relabelling. 
The same holds for $\pi^+ \pi^+ K^+$ and $\pi^+ K^+ K^+$ scattering. 
Thus we can treat both systems simultaneously without specifying 
the choice of $(m_1,m_2)$.

We find that the contribution from diagram (a) is given by
\begin{align}
\cM_3^{(a)} = - \cM_2^{(2)\text{,off1}}(p_1,p_{1'}) \frac{1}{b^2_{(a)} - m_1^2 + i\epsilon}  
\cM_2^{(1)\text{,off1}}(k_{1'},k_2), \quad b_{(a)}=p_1+p_{1'}-k_1,
\end{align}
where the off-shell two-particle amplitudes are given by
\begin{align}
\begin{split}
\cM_2^{(2)\text{,off1}}(p_1,p_{1'}) =& -\frac{2 p_1\cdot p_{1'}}{F^2} 
+ \frac{1}{3 F^2}(b_{(a)}^2 - m_1^2)\,, 
\\
\cM_2^{(1)\text{,off1}}(k_{1'},k_2) =& -\frac{ k_{1'}\cdot k_2}{F^2} 
+ \frac{1}{6 F^2}(b_{(a)}^2 - m_1^2)\,. 
\label{eq:M2off1}
\end{split}
\end{align}
The notation ``off1'' indicates that it is a particle of flavor 1 that is off shell.
For diagram (b) the result is similar,
\begin{equation}
 \cM_3^{(b)} = - \cM_2^{(1)\text{,off2}}(p_{1'},p_2) \frac{1}{b^2_{(b)} - m_2^2+ i\epsilon}
   \cM_2^{(1)\text{,off2}}(k_{1'},k_2)\,, \quad b_{(b)}=p_{1'}+p_2-k_1\,,
\end{equation}
where 
\begin{align}
\cM_2^{(1)\text{,off2}}(k_{1'},k_2) =& -\frac{ k_{1'}\cdot k_2}{F^2} 
+ \frac{1}{6 F^2}(b_{(b)}^2 - m_2^2)\,,
\label{eq:M2off2}
\end{align}
with the particle of flavor 2 being off shell.

As noted above, the full contribution of the OPE diagrams to $\cM_3^{\rm LO}$ requires
the addition of other terms.
For diagram (a) one adds the result of interchanging $k_1$ and $k_{1'}$, 
and for both resulting diagrams one adds the result of a PT transformation,
which is obtained by making the interchanges $k_i \leftrightarrow p_i$.
For diagram (b), one interchanges flavors $1$ and $1'$ for both initial and final states.
We hold off on adding these other contributions until we have made the subtractions.

Finally, for the diagram with the six-point vertex we find
\begin{equation}
F^4 \cM_3^{(c)} = \frac{1}{3} M^2 \Delta - \frac{1}{36}  M^2\Delta_2^S +  \frac{1}{12}  M^2 \wt t_{22} + \frac{2}{3}\left(2 m_1 m_2 + m_1^2\right)\,, 
\label{eq:Mdf3c}
\end{equation}
where we are using the kinematic quantities defined in \Cref{eq:kindefs}.
We stress that no subtraction is needed for this contribution.

\subsection{Subtraction terms and $\cM_{\df,3}$}

We next evaluate $\cD^{\rm LO}$, which is given in \Cref{eq:subLO}.
The terms on the first two lines of this result subtract the contributions from OPE diagrams
of type \Cref{fig:diagrams}(a), while those on the third line contain the subtractions
for diagrams of type \Cref{fig:diagrams}(b).
We stress that the subtraction can be done diagram by diagram.
The results for the subtraction terms are very similar to those for the original diagrams,
except that the off-shell two-particle amplitudes are replaced by their on-shell correspondents.
Thus, for example, the subtraction term for \Cref{fig:diagrams}(a) is
\begin{equation}
\cD^{(a)} = -\cM_2^{(2)\text{,on}}(p_1,p_{1'}) \frac{1}{b_{(a)}^2 - m_1^2+ i\epsilon} 
 \cM_2^{(1)\text{,on}}(k_{1'},k_{2})\,,
\end{equation}
where 
\begin{align}
\begin{split}
\cM_2^{(2)\text{,on}}(p_1,p_{1'}) = -\frac{2 p_1 \cdot p_{1'}}{F^2}, \quad
\cM_2^{(1)\text{,on}}(k_{1'},k_2) = -\frac{ k_{1'}\cdot k_2}{F^2}\,. 
\label{eq:M2on}
\end{split}
\end{align}
Similarly, for diagram (b) one simply drops the $b_{(b)}^2-m_2^2$ contributions in
the expression for $\cM_2^{(1),\rm off2}$, \Cref{eq:M2off2}.

Using these results, we obtain the divergence-free matrix elements
\begin{align}
F^4 \cM_\text{df,3}^{(a)} &= 
\frac{1}{3}(p_1\cdot p_{1'}+k_{1'}\cdot k_2) - \frac{1}{18}(b_{(a)}^2 - m_1^2)\,,
\\
F^4 \cM_\text{df,3}^{(b)} &= 
\frac{1}{6}(p_{1'}\cdot p_2+k_{1'}\cdot k_2) - \frac{1}{36}(b_{(b)}^2 - m_2^2)\,.
\end{align}
If we now add to these the above-described three additional contributions for each type of diagram,
we find
\begin{align}
F^4 \cM_\text{df,3}^{(a),\text{all}} &= 
\frac{M^2}{18} \left( 6 \Delta +3 \Delta_2^S - 2 \wt t_{22} \right) + \frac{4}{3} (m_1m_2+m_1^2)\,, 
\label{eq:Mdf3a}
\\ 
F^4 \cM_\text{df,3}^{(b),\text{all}} &= 
\frac{M^2}{36} \left( 12 \Delta -5\Delta_2^S + \wt t_{22} \right) + \frac{4}{3} m_1 m_2\,.
\label{eq:Mdf3b}
\end{align}

\subsection{Final result}

At this point we note that the contribution from each of the three diagrams,
i.e.~those from \Cref{eq:Mdf3c,eq:Mdf3a,eq:Mdf3b},
has the form expected, based on symmetries, given in \Cref{eq:thexp}.
There are isotropic terms with either no dependence or linear dependence on
$s$ (or, equivalently, on $\Delta$), together with nonisotropic
$\Delta_2^S$ and $\wt t_{22}$ terms.
We note that working to LO in ChPT leads to contributions with up to two powers of momenta,
which are thus at most linear in the Mandelstam variables.
Thus it corresponds to working to linear order in the threshold expansion described
in \Cref{sec:Kdf}.

However, when we combine the three contributions to get the final result,
we find that the nonisotropic terms cancel
\begin{align}
F^4 \cK_{\df,3}^{\rm LO} 
&= 
F^4 \left( \cM_\text{df,3}^{(a),\text{all}}+\cM_\text{df,3}^{(b),\text{all}} + \cM_3^{(c)} \right)
\\
&= M^2 \Delta + 4 m_1 m_2 + 2 m_1^2\,. 
\label{eq:KdfLOres}
\end{align}
We do not have an explanation for this cancellation.
It implies that the contributions to the coefficients of the nonisotropic terms,
i.e.~$\cK_{\df,3}^{B,1}$ and $\cK_{\df,3}^{E,1}$ in \Cref{eq:thexp},
 can appear first at NLO in ChPT.
Thus while $\cK_{\df,3}^{\iso,0}\sim \cK_{\df,3}^{\iso,1} \sim m^2/F^4$,
we expect that
$\cK_{\df,3}^{B,1}\sim\cK_{\df,3}^{E,1}\sim m^4/F^6$,
where $m$ is a generic meson mass.

\section{Numerical applications}
\label{sec:numerical}

In this section we provide two numerical applications of the quantization conditions that
we have implemented. 
First, we compare the energy of the ground state
of a completely nondegenerate system with the $1/L$ expansion derived in \Cref{app:threshold}. 
Our aims here are to provide a check of our numerical implementation (which must agree increasingly
well with the truncated $1/L$ expansion as $L$ increases)
and to see how rapidly the $1/L$ expansion converges.
We choose the nondegenerate system both to advertise that the code for this is available,
and also because the threshold expansion for this system has not previously been derived.

In our second example, we use the $2+1$ quantization condition to 
predict the energy shifts for several levels in the $\pi\pi K$ and $K K \pi$ systems, 
choosing parameters that are likely to be used in near-term lattice simulations.
Our main aim here is to illustrate the precision needed to determine the different components
of $\cK_2$ and $\Kdf$.

\subsection{Testing the threshold expansion}
\label{sec:thrnum}


The expansion of the energy of the ground state when the total momentum vanishes is
usually referred to as the ``threshold expansion.''
In \Cref{app:threshold} we obtain the following result for the threshold expansion for
three nondegenerate scalars,
\begin{align}
\begin{split}
 L^3 \Delta E &= \sum_{i=1}^3 \frac{2 \pi a_0^{(i)}}{\mu_i} \left[1 - \frac{a_0^{(i)}}{\pi L} \mathcal I 
  + \left(\frac{a_0^{(i)}}{\pi L} \right)^2 (\mathcal I ^2 - \mathcal{J})  
  + \frac{a^{(j)}_0 a^{(k)}_0}{(\pi L)^2} 2\mathcal{J} \right] + O(L^{-6})\,,
\end{split}
  \label{eq:NNLOthreshold}
\end{align}
where $i$, $j$, and $k$ are ordered cyclically, $\mu_i$ is the reduced mass of the $jk$ system,
and $a^{(k)}_0$ is the scattering length of particles $i$ and $j$. 

To test our implementation of the nondegenerate quantization condition,
we choose the following mass ratios 
\begin{equation}
\frac{m_2}{m_1} = 1.5, \quad \frac{m_3}{m_1} = 0.5 \, 	.
\label{eq:NDmassratios}
\end{equation}
We express all quantities in units of the mass of the first particle, $m_1$, 
including the size of the box $m_1 L$.
Since the effective range does not enter the threshold expansion until $O(1/L^6)$,
we keep only the leading term in the effective-range expansion for the phase shift
\begin{equation}
q \cot \delta^{(ij)}_0 = -\frac{1}{a_0^{(k)}}.
\end{equation}
Similarly, since neither $p$-wave two-particle interactions nor $\Kdf$ enters into the threshold
expansion until higher order, we consider the quantization condition with only $s$ waves and
with $\Kdf=0$. We choose the following values for the scattering lengths,
\begin{equation}
m_1 a_0^{(1)} = 0.7, \quad m_1 a_0^{(2)} = 0.5, \quad m_1 a_0^{(3)} = 0.3.
\label{eq:NDscatt}
\end{equation}
Our choices in \Cref{eq:NDmassratios,eq:NDscatt} imply significant
nondegeneracy in masses, and symmetry breaking in interactions. In this way we are performing
a fairly robust test of our implementation.

\begin{figure}[h!]
  \centering  
\begin{subfigure}{.49\textwidth}
  \centering
  \includegraphics[width=1\linewidth]{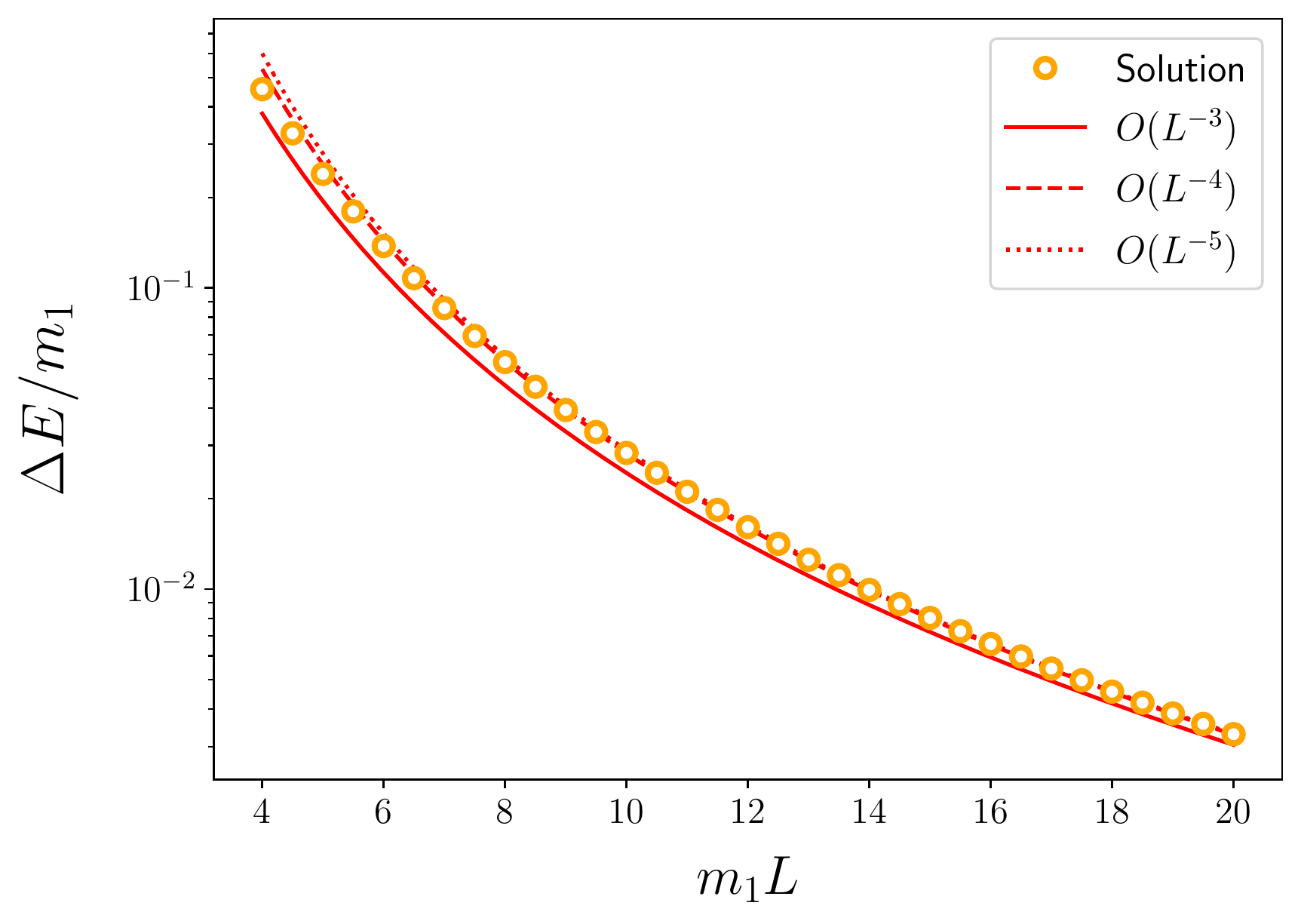}  
  \caption{}
  \label{fig:th1}
\end{subfigure}
\begin{subfigure}{.49\textwidth}
  \centering
  \includegraphics[width=1\linewidth]{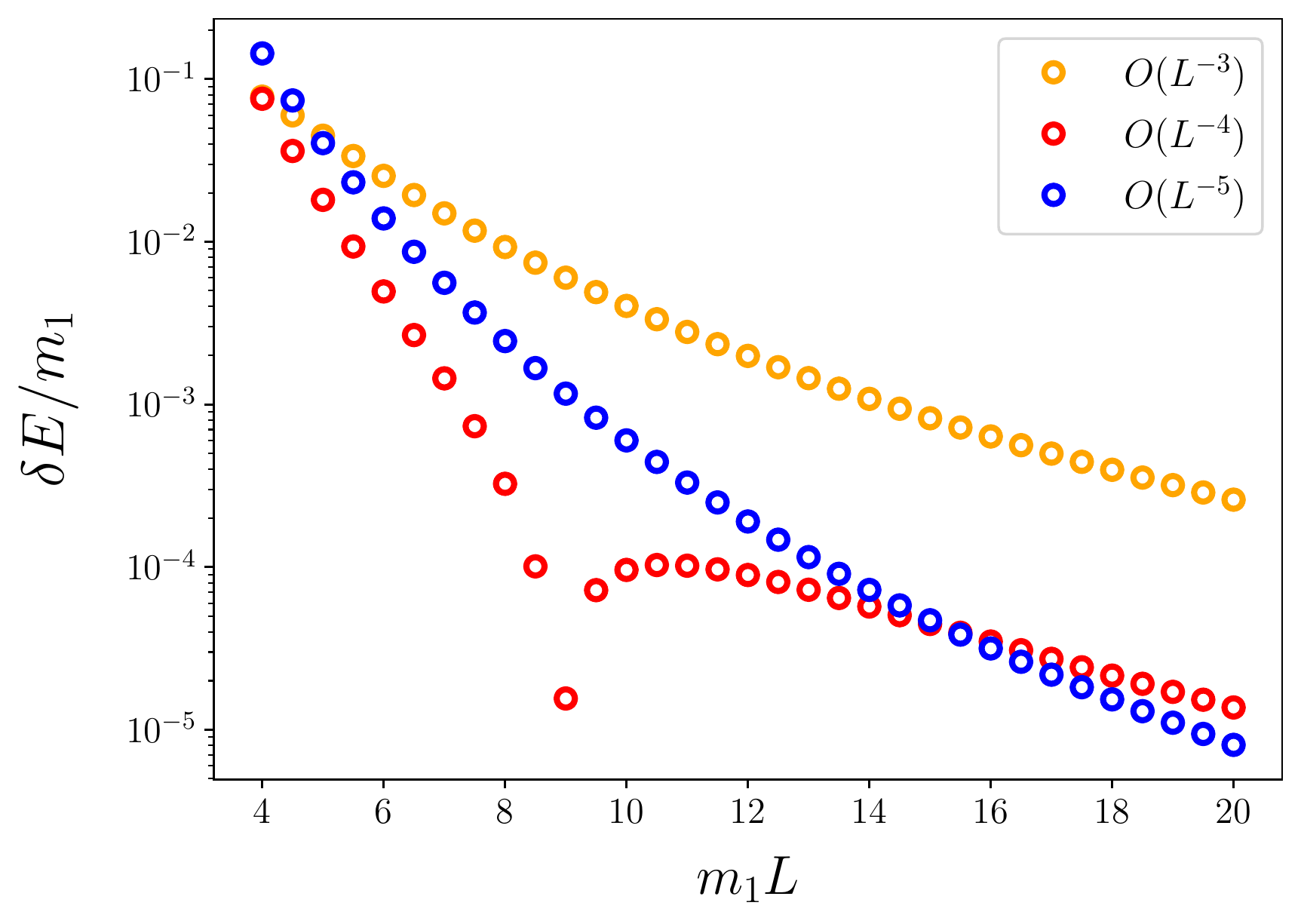}  
  \caption{}
  \label{fig:th2}
\end{subfigure}
\caption{Numerical comparison between the results from the quantization condition and the threshold expansion. (a) Energy shift as a function of the box size.
Orange dots show the numerical solutions to the quantization condition, 
while the various red lines are the theoretical predictions of \Cref{eq:NNLOthreshold}, 
truncated to different orders in the $1/L$ expansion. 
(b) Box-size dependence of the quantity in \Cref{eq:difference},
which is the absolute value of the difference between the threshold expansion at various orders
and the numerical solution as a function of $m_1 L$. 
}
\label{fig:th}
\end{figure}

The numerical results are shown in \Cref{fig:th}. 
The left plot shows the dependence of the energy shift on the size of the box.  
We observe that, while the leading-order result captures the overall behavior of the exact results,
albeit with a noticeable offset,
adding in the $O(L^{-4})$ and $O(L^{-5})$ terms leads to a much closer agreement.
To study this more quantitatively, 
we show in the right plot the difference between the numerical and analytical results,
\begin{equation}
\delta E \equiv | \left( \Delta E \right)^{\rm QC3} - \left( \Delta E\right)^{\rm th}|.
\label{eq:difference}
\end{equation}
For small $m_1 L$ it appears that the threshold expansion truncated at the $L^{-4}$ term does
better than that including the $L^{-5}$ term, indicating a breakdown in convergence.
However, at the largest volumes shown the expected ordering of curves sets in.
This gives us confidence in our implementation, and also indicates that relatively large volumes
are needed for the threshold expansion to give a good approximation to the energy shift.

\subsection{Model results for the $\pi^+\pi^+ K^+$ and $\pi^+K^+ K^+$ energy levels}
\label{sec:numericaltoy}
 
A major motivation for developing the formalism for $2+1$ systems
was that extensive lattice results for such systems will be available soon.
Specifically, recent lattice calculations of multiple energy levels for
$3\pi^+$ and $3K^+$ systems~\cite{Horz:2019rrn,Culver:2019vvu,Fischer:2020jzp,Hansen:2020otl,
Alexandru:2020xqf,Brett:2021wyd,Blanton:2021llb}
can be relatively straightforwardly generalized to 
$\pi^+\pi^+K^+$ and $\pi^+K^+K^+$.
The only previous study of the latter systems of which we are aware considered the threshold 
states alone~\cite{Detmold:2011kw}.

An important consideration when fitting the quantization condition to results for the spectrum
is the precision needed from the lattice calculation in order to determine the various parameters
that enter $\cK_2$ and $\Kdf$. Here we give an indication of the required precision for
$\pi^+\pi^+K^+$ and $\pi^+K^+K^+$ systems. Specifically, we determine the energy
shift for several levels in the rest frame and a moving frame. These levels are in
several irreps of the cubic group (which is the symmetry group of the cubic box that we consider).
We choose the $\pi^+$ and $K^+$ masses from the N203 ensemble~\cite{Bruno:2016plf} 
created by  the Coordinated Lattice Simulations (CLS) effort,
which is one of the ensembles used in the recent detailed analysis of $3\pi^+$ and
$3K^+$ systems in Ref.~\cite{Blanton:2021llb}. 
The parameters that we need are
\begin{equation}
M_K/M_\pi = 1.278, \quad M_\pi/F_\pi=  3.433, \quad M_K/F_K=4.153,\quad  M_\pi L = 5.4053, \label{eq:paramN203}
\end{equation}
where $M_\pi$ and $F_\pi$ are the pion mass and decay constant, and $M_K$ and $F_K$ the analogous quantities for the kaon.  

To make predictions for the energy shifts, we set $\cK_2$ and $\Kdf$ to their LO expressions
in SU(3) ChPT,
which we expect will be a reasonable approximation for levels that are not at too high energies.
This implies that all interactions are purely $s$-wave. 
The $\pi^+\pi^+$ and $K^+K^+$ phase shifts are given by
\begin{align}
\frac{q}{M_\pi} \cot \delta^{\pi\pi}_0 &= \frac{M_\pi \sqrt{s} }{s - 2 M_\pi^2} \left( -\frac{16 \pi F_\pi^2}{M_\pi^2} \right), \\
\frac{q}{M_K} \cot \delta^{KK}_0 &= \frac{M_K \sqrt{s} }{s - 2 M_K^2} \left(- \frac{16 \pi F_K^2}{M_K^2} \right)\,,
\end{align}
where $s$ is the two-particle squared total four-momentum.
At LO in ChPT $F=F_\pi=F_K$, and we have chosen the decay constant corresponding to the particles
that are scattering.
For $\pi^+K^+$ scattering the LO expression can be written
\begin{equation}
\frac{q}{M_\pi} \cot \delta^{\pi K}_0 = \frac{M_\pi \sqrt{s} }{s-M_\pi^2 - M_K^2} \left( -\frac{16 \pi F_\pi^2}{M_\pi^2} \right)\,,
\end{equation}
which we use in the $\pi^+\pi^+ K^+$ case,
or
\begin{equation}
\frac{q}{M_K} \cot \delta^{\pi K}_0 = \frac{M_K \sqrt{s} }{s-M_\pi^2 - M_K^2} \left(- \frac{16 \pi F_K^2}{M_K^2} \right)\,,
\label{eq:piKswave}
\end{equation}
which we use for $\pi^+ K^+ K^+$. 
Note that the choice of $M_\pi$ or $M_K$ to make quantities
dimensionless is arbitrary, since the overall factors of mass cancel. 
We also note that the cutoff function described in \Cref{sec:cutoff} implies that we 
will evaluate $\delta^{\pi K}_0$ down to $s_\text{min}= M_K^2 - M_\pi^2$, 
where $q^2 = - M_\pi^2$. By contrast, $\delta^{\pi\pi}_0$ and $\delta^{KK}_0$ 
will be evaluated down to $s_\text{min}=0$, where $q^2 = - M_\pi^2$ and $q^2 = - M_K^2$, respectively.

For $\Kdf$ we use the results of Sec.~\ref{sec:chpt}. Again we have to choose which
decay constants to use in the LO expressions, and our approach is
to divide each particle mass by the corresponding decay constant.
In this way we obtain from \Cref{eq:KdfLOres} the following LO results
\begin{align}
M_\pi^2\cK_\text{df,3}^{\pi\pi K} &=  2\frac{M_\pi^4}{F_\pi^4} + 4\frac{M_\pi^3}{F_\pi^3}\frac{M_K}{F_K} + \frac{M_\pi^2}{F_\pi^2} \left(  2\frac{M_\pi}{F_\pi}+\frac{M_K}{F_K}  \right)^2 \Delta\,,
\\
M_K^2\cK_\text{df,3}^{KK\pi} &=  2\frac{M_K^4}{F_K^4} + 4\frac{M_K^3}{F_K^3}\frac{M_\pi}{F_\pi} + \frac{M_K^2}{F_K^2} \left(  2\frac{M_K}{F_K}+\frac{M_\pi}{F_\pi}  \right)^2 \Delta \,.
\end{align}

\begin{figure}[h!]
  \centering  
\begin{subfigure}{.49\textwidth}
  \centering
  \includegraphics[width=1\linewidth]{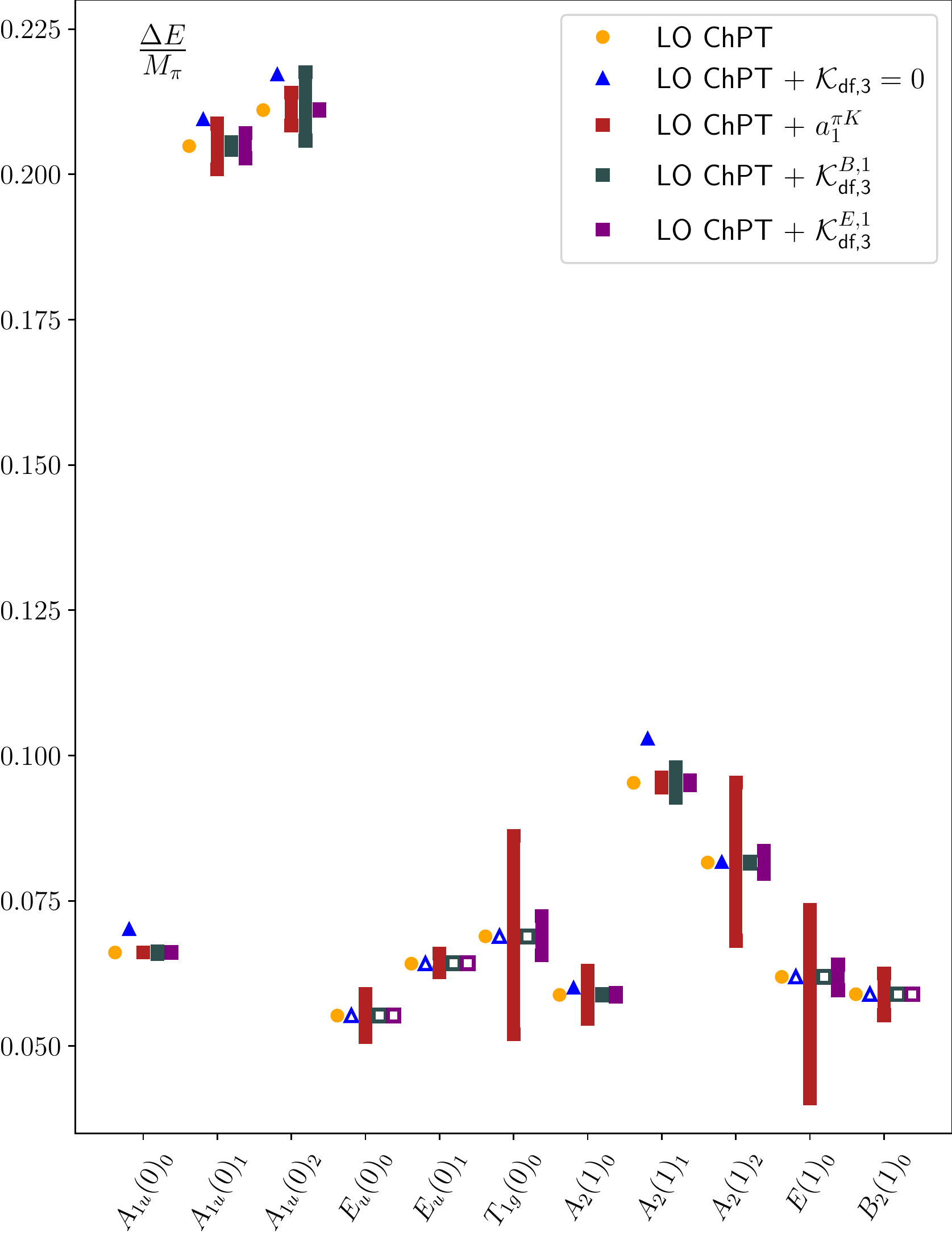}  
  \caption{}
\end{subfigure}
\begin{subfigure}{.49\textwidth}
  \centering
  \includegraphics[width=1\linewidth]{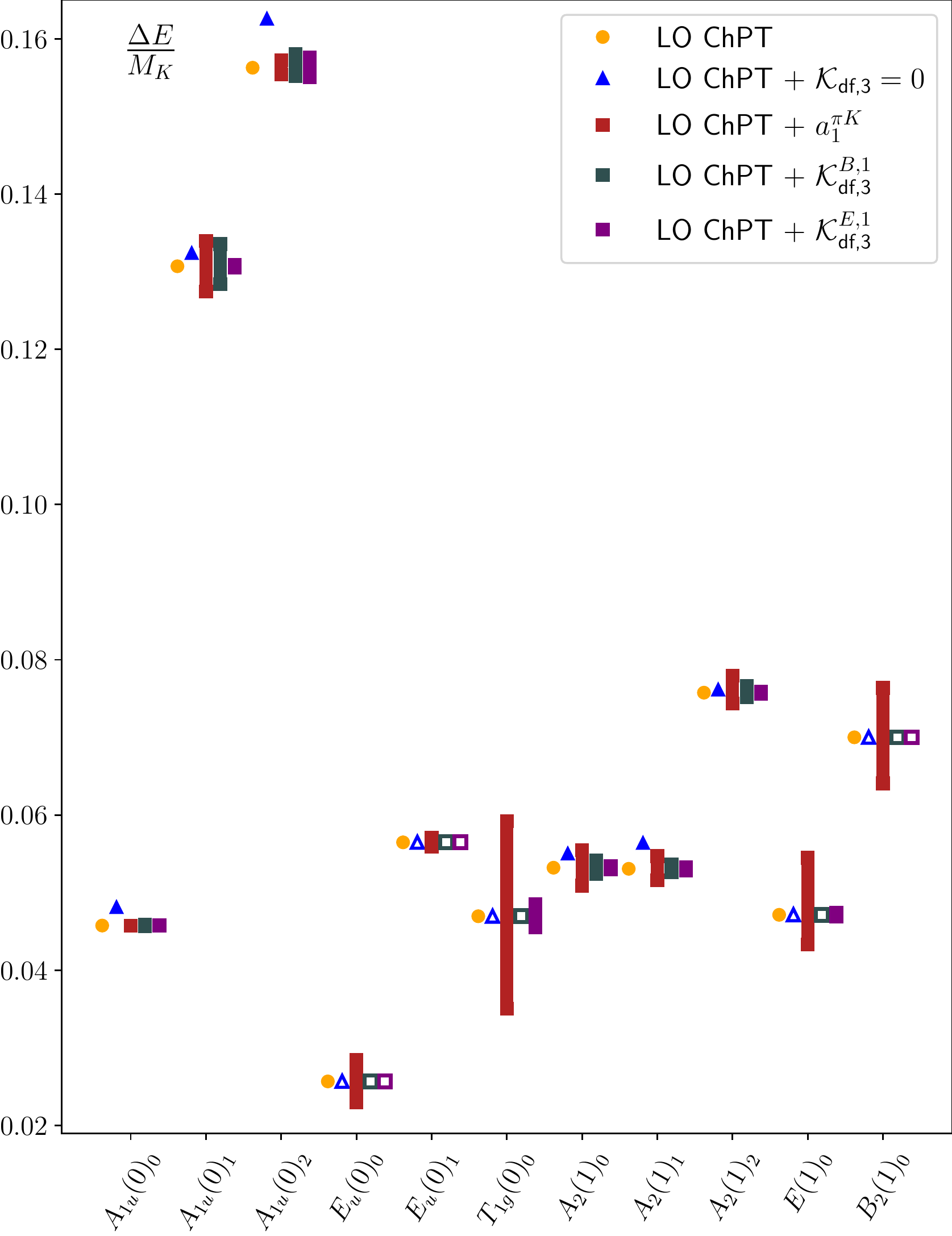}  
  \caption{}
\end{subfigure}
\caption{Shifts in the CMF energy of various levels with different choices of parameters in
$\cK_2$ and $\Kdf$ for (a) the $\pi^+\pi^+K^+$ and (b) the $\pi^+K^+K^+$ systems.
The levels are denoted along the bottom of the figures using the notation
${\rm irrep}(\bm{d}_{\rm ref}^2)_{\rm level}$, where $\bm d_{\rm ref}=\bm P L/(2\pi)$ describes
the frame, while the subscript $0,1,\dots$ denotes the level, with $0$ being the ground state
in the given irrep and frame. For each level, there are five points, as discussed in the text.}
\label{fig:spectrum}
\end{figure}

Using these inputs we then numerically solve the quantization condition and determine the
shifts of the CMF energy $E^*= \sqrt{E^2- \bm P^2}$
for each level from the corresponding free energy.
These shifts are plotted for the several
levels in \Cref{fig:spectrum} as the points denoted ``LO ChPT.''
These are all the levels that appear in the
lowest three orbits in the rest frame and the first moving frame. 
For reference, the CMF energies of
these levels are collected in \Cref{tab:spectrum}.

\begin{table}[b]
\centering
\begin{tabular}{c|c|c|c|c}
{$\bm{d}_{\rm ref}$}&{$[d_{1}^2,d_{1'}^2,d_{2}^2]$}& irreps 
&$E^{*\rm free}_{\pi\pi K}/M_\pi$ &$E^{*\rm free}_{\pi K K}/M_K$ \\%
\hline%
(0, 0, 0)&{[}0, 0, 0{]}&$A_{1u}$ & 3.278 & 2.782 \\%
&{[}0, 1, 1{]}& $A_{1u} \oplus E_u \oplus T_{1g}$ & 4.260 & 3.486 \\%
&{[}1, 1, 0{]}& $E_u \oplus T_{1g}$ & 4.345 & 3.552\\%
\hline%
(0, 0, 1)
&{[}0, 0, 1{]}& $A_2$ & 3.542 & 2.999 \\%
&{[}0, 1, 0{]}& $A_2$ & 3.630 & 3.068\\%
&{[}0, 1, 2{]}& $A_2 \oplus B_2 \oplus E$ & 4.467 & 3.652\\%
\end{tabular}%
\caption{\label{tab:spectrum} 
Noninteracting CMF energies,
$E^{*\rm free} = \sqrt{E_{\rm free}^2 - \bm P^2}$,
and little group irreps,
for the $\pi^+\pi^+K^+$  and $\pi^+ K^+ K^+$ levels appearing in \Cref{fig:spectrum}.
Frames are denoted by $\bm{d}_{\rm ref} = {\bm P} L/(2\pi)$.
Orbits are specified by the squared momenta of the three particles, 
e.g.~$d_1^2= \bm p_1^2 (L/(2\pi))^2$.
The flavor labels are $\{1,1',2\}=\{\pi^+,\pi^+,K^+\}$ for $\pi^+\pi^+K^+$,
and $\{1,1',2\}=\{K^+,K^+,\pi^+\}$ for $\pi^+ K^+ K^+$.
}
\end{table}

Since leading-order two-particle interactions give the dominant contribution to the energy shifts, 
an important question is how large are the effects of higher-order two-
and three-particle interactions on the energy shifts.
To answer this question, we include, for each level, four more points in \Cref{fig:spectrum}.
The first shows the effect of setting $\Kdf=0$ while keeping the $s$-wave phase shifts
unchanged. We observe that some levels do not depend at all on the isotropic $\Kdf$ predicted by
LO ChPT, whereas for the others there is a shift. To determine the isotropic component
of $\Kdf$ one needs to determine the energy shifts with errors no larger than $\sim 5\%$.
This precision has been achieved for some levels
in the $3\pi^+$ and $3K^+$ spectra of Ref.~\cite{Blanton:2021llb}.

The remaining three points for each level in \Cref{fig:spectrum} display the impact of terms
of higher order in ChPT. The first considers the $p$-wave $\pi^+ K^+$ interaction, which
appears at NLO in ChPT. To estimate the size of the
$p$-wave scattering length (defined in \Cref{eq:EREp})
we use the NLO SU(3) ChPT results from Ref.~\cite{Bernard:1990kw} from which it follows that 
\begin{equation}
M a_1^{\pi K} \sim \frac1{16\pi F^2} \frac{M^2}{(4\pi F)^2}\,,
\label{eq:a1piK} 
\end{equation}
where the unknown overall constant of $\cO(1)$ arises from a combination of loops
and low-energy coefficients, and $M$ represents a combination of pion and kaon masses
that varies from term to term.
To implement this in practice we choose 
\begin{equation}
M_\pi^3 a_1^{\pi K} = c \frac{M_\pi+M_K}{2M_\pi M_K}
\frac{M_\pi^2}{(4\pi F_\pi)^2} \left(\frac{M_\pi^2}{(4\pi F_\pi)^2} + \frac{M_K^2}{(4\pi F_K)^2}\right)\,,
\end{equation}
for the $\pi^+\pi^+ K^+$ system, and
\begin{equation}
M_K^3 a_1^{\pi K} = c \frac{M_\pi+M_K}{2M_\pi M_K}
\frac{M_K^2}{(4\pi F_K)^2} \left(\frac{M_\pi^2}{(4\pi F_\pi)^2} + \frac{M_K^2}{(4\pi F_K)^2}\right)\,,
\end{equation}
for the $\pi^+K^+ K^+$ system. We vary $c$ in the range $-3 \le c \le 3$,
which corresponds, after keeping track of factors of $\pi$, to the overall constant in
\Cref{eq:a1piK} being $\sim 1$.
This leads to the bands shown as the third (maroon) entry for each level in \Cref{fig:spectrum}.
We note that all levels are sensitive to $a_1^{\pi K}$, and that for most, though not all, the
sensitivity is greater than that to the LO part of $\Kdf$. 

Finally we consider the NLO contributions to $\Kdf$. Both $\cK_{\df,3}^{B,1}$ and 
$\cK_{\df,3}^{E,1}$ appear at this order;
we choose
\begin{equation}
M_\pi^2 \KB = c_B \frac{M_\pi^4}{F_\pi^4} \frac{M_K^2}{F_K^2}\,,
\qquad
M_\pi^2 \KE = c_E \frac{M_\pi^4}{F_\pi^4} \frac{M_K^2}{F_K^2}
\end{equation}
for the $\pi^+\pi^+K^+$ system,
and 
\begin{equation}
M_K^2 \KB = c_B \frac{M_\pi^2}{F_\pi^2} \frac{M_K^4}{F_K^4}\,,
\qquad
M_K^2 \KE = c_E \frac{M_\pi^2}{F_\pi^2} \frac{M_K^4}{F_K^4}
\end{equation}
for the $\pi^+ K^+K^+$ system.
We then vary the constants in the range $-1 \le c_B,c_E \le 1$.
Given that the standard estimate of NLO effects involves a loop factor of $1/(4\pi)^2$,
which we are not including here, these are very aggressive ranges. We justify them by the
observation that, if we were to use the analogous chiral estimates for the $d$-wave
contributions to $\Kdf$ in the $3\pi^+$ and $3 K^+$ systems, then to match the values found
in Ref.~\cite{Blanton:2021llb}, 
a similarly aggressive estimate would be needed. We also note that the
scattering amplitude $\cM_3$ does not diverge when $\Kdf\to \pm \infty$.

The final two points for each level in \Cref{fig:spectrum} show the effect of making separate
variations as just described in $\KB$ and $\KE$. We observe that $\KB$ contributes only
to levels in the trivial irreps, but can contribute as much as the isotropic parts of $\Kdf$ and
$a_1^{\pi K}$ to certain levels. $\KE$ does contribute to some of the levels in nontrivial irreps,
which might make it somewhat simpler to determine than $\KB$.

\section{Conclusions}
\label{sec:conc}
%

The goal of this paper is to prepare for the analysis of lattice QCD data for systems 
involving nondegenerate particles. 
For this, we have implemented the finite-volume formalism derived in 
Refs.~\cite{\BSnondegen,\BStwoplusone}, i.e.,
for both fully nondegenerate and $2+1$ systems. 
We have made our code public in the following repository~\cite{coderepo}. 
The description of the code, together with some examples, can be found in \Cref{app:code}.

While many details are similar to the case of identical particles, 
several additional technical features arise with nondegenerate particles.
These have been discussed in \Cref{sec:implement},
and include the definition of the sum-minus-integral difference  for nonidentical particles, 
as well as the need for a modified cutoff function. 
The latter is necessary to avoid the $t$-channel cut, 
which is displaced with respect to the case of identical particles. 
Furthermore, odd angular momenta appear when the particles are nondegenerate, 
and the extension of the group theory to handle $p$-wave interactions
is explained in \Cref{sec:projections}.

When implementing the quantization conditions, it is useful to have analytic checks in certain
limits. One such check is provided by the expansion of the
energy of the three-particle ground state in powers of $1/L$. This has been worked out previously
for the identical-particle and $2+1$ systems. Here, in \Cref{app:threshold}, we extend the derivation
to the fully nondegenerate case, including terms up to $O(L^{-5})$.
We show an example of this check on the implementation for a completely nondegenerate system
in \Cref{sec:thrnum}.

As noted in the Introduction, the immediate application for this formalism is to systems
with a mix of light pseudoscalar mesons, in particular $\pi^+ \pi^+ K^+$ and $\pi^+ K^+ K^+$.
Thus we have focused much  of the discussion on such $2+1$ systems.
To study them in practice, a parametrization of $\Kdf$ is needed. 
This can be achieved using a systematic expansion about threshold that
accounts for all symmetries~\cite{\dwave},
and if we work to linear order in this expansion, then only four terms contribute~\cite{\BStwoplusone}.
Two of these involve only two-particle $s$ waves, while the other two also include $p$ waves.
Due to the possibility of having different flavors  of spectator particles, 
the decomposition of $\Kdf$ into the finite-volume kinematic variables becomes more complicated
than that for identical particles, and we have worked this out explicitly in \Cref{sec:Kdf}.
The code for $2+1$ systems that we provide includes this implementation.

The parameters in this threshold expansion of $\Kdf$ can, in principle, be predicted in ChPT,
with the LO calculation being relatively straightforward.
The results can serve as a guide for what to expect when fitting to results from lattice simulations.
Thus we have extended the work of Ref.~\cite{Blanton:2019vdk}, in which the LO
prediction was worked out for $3\pi^+$ scattering,
to the  $\pi^+ \pi^+ K^+$ and $\pi^+ K^+ K^+ $ systems.
The results, provided in \Cref{sec:chpt}, turn out to be completely isotropic, 
i.e., both two- and three-particle interactions involve only $s$ waves.
Thus two of the four parameters in the threshold expansion are predicted to vanish
at LO in ChPT, although all four terms are present at intermediate stages.

The finite-volume spectrum of three particles depends primarily on  two-particle interactions,
with a subleading contribution from $\Kdf$. It is important to understand how large the various
contributions are to the spectral levels, in order to determine how precisely one must
determine their energies in a simulation.
To investigate this, in \Cref{sec:numericaltoy}, we explore the impact of
the different choices for the two- and three-particle K matrices on
selected levels in the $\pi^+ \pi^+ K^+$ and $\pi^+ K^+ K^+$ spectra. 
We use the values of the masses, decay constants, and box size that match those of the
N203 CLS ensemble~\cite{Bruno:2016plf}.
We use both the LO ChPT predictions, and estimates of NLO contributions based on
chiral power counting.
The results emphasize the importance of using several frames and levels in all available irreps, in order to
determine the two-particle $p$-wave interaction and the nonisotropic terms in $\Kdf$.

Finally, we recall the recent analysis  in Ref.~\cite{Blanton:2021llb}
of $3\pi^+$ and $3K^+$ systems using hundreds of energy levels in total. 
It appears that extending this work to mixed systems of $K^+$ and $\pi^+$ mesons
is technically feasible. Fitting the results to the $2+1$ quantization condition that we
have implemented here should allow a determination
of both the $I=3/2$ $\pi K$ scattering amplitudes in $s$ and $p$ waves 
as well as some of the terms in $\Kdf$. 
Some new technical challenges will need to be faced, 
e.g., how to simultaneously fit the spectra of multiple three- and two-particle systems,
in this case $3\pi^+$, $\pi^+ \pi^+ K^+$, $\pi^+ K^+ K^+$, and $3 K^+$,
as well as $\pi^+\pi^+$, $\pi^+ K^+$, and $K^+ K^+$.
An important challenge will be the reliable inclusion of correlations between
such a large number of levels.

\acknowledgments

We thank Will Detmold and Drew Hanlon for useful discussions.

TDB is supported in part by the United States Department of Energy (USDOE) under contract No.~DE-FG02-93ER-40762, and also by USDOE Grant No.~DE-SC0021143.
FRL acknowledges the support provided by the European project H2020-MSCA-ITN-2019//860881-HIDDeN, the Spanish project FPA2017-85985-P, and the Generalitat Valenciana grant PROMETEO/2019/083. FRL has also received financial support from Generalitat Valenciana through the plan GenT program (CIDEGENT/2019/040). The work of FRL has been supported in part by the U.S.~Department of Energy, Office of Science, Office of Nuclear Physics, under grant Contract Numbers DE-SC0011090 and DE-SC0021006. The work of SRS is supported in part by the USDOE grant No.~DE-SC0011637.

\clearpage
\appendix

\section{Further details of implementation} 
\label{app:details}


In this appendix we provide further technical details of the implementation of the
quantization condition for $2+1$ systems, which has been described in Sec.~\ref{sec:implement}.

\subsection{Real spherical harmonics}

As noted in the main text, in practice we use the real version of spherical harmonics, following
Ref.~\cite{\dwave}. In this way we do not need to keep track of which harmonics to complex conjugate,
and all matrices appearing in the quantization condition are real and symmetric.
The real spherical harmonics satisfy the same orthonormality conditions as the more standard
complex version.
In this work we only need those for $\ell = 1$, as well as the standard form for $\ell=0$.
The former are
\begin{equation}
pY_{11}(\hat p) = \sqrt{\frac3{4\pi}} p_x\,,\quad
pY_{10}(\hat p) = \sqrt{\frac3{4\pi}} p_z\,,\quad
pY_{1-1}(\hat p) = \sqrt{\frac3{4\pi}} p_y\,.
\end{equation}

\subsection{Evaluating $\wt F^{(i)}$}

To evaluate $\wt F^{(i)}$, \Cref{eq:Ft}, we use the UV regularization introduced
in Ref.~\cite{\KSS}.
Dropping exponentially suppressed terms,
\Cref{eq:Ft} can be brought into the form 
\begin{align}
\left[\wt F^{(i)}\right]_{p' \ell' m';p \ell m} &=
 \frac{\delta_{\bm p' \bm p}H^{(i)}(\bm p)}{16 \pi^2 L^4 \omega_p^{(i)} (E-\omega_p^{(i)})}
 \left[ \sum_{\bm n_a} - {\rm PV}\!\! \int d^3 n_a \right]
 \frac{e^{\alpha(x^2-r^2)}}{x^2-r^2}
 \frac{\cY_{\ell' m'}(\bm r) \cY_{\ell m}(\bm r)}{x^{\ell'+\ell}}\,,
 \label{eq:Ftb}
 \end{align}
 where $\bm a = \bm n_a (2\pi/L)$, $x=q_{2,p}^{*(i)} L/(2\pi)$, and
 $\bm r$ is a vector given in terms of ${\bm n_{pP}=(\bm p - \bm P) L/(2\pi)}$ by
 \begin{equation}
 \bm r = \bm n_a + \bm n_{pP} \left[
\frac{ \bm n_a \cdot \bm n_{pP}}{\bm n_{pP}^2}
\left(\frac1{\gamma}-1\right) + \frac{\xi}{\gamma}\right]\,,
  \end{equation}
 where
 \begin{equation}
 \xi = \frac12 \left( 1 + \frac{m_1^2-m_2^2}{\sigma_i}\right)\,,
 \end{equation}
 and
 \begin{equation}
 \gamma = \frac{E-\omega_p^{(i)}}{\sqrt{\sigma_i}}\,.
 \end{equation}
  Ultraviolet regularization is provided by $\alpha > 0$~\cite{\KSS}, with the dependence on
 $\alpha$ being exponentially suppressed in $L$.
The result (\ref{eq:Ftb}) agrees with the two-particle nondegenerate result given in
Refs.~\cite{Briceno:2012yi,Davoudi:2011md,Fu:2011xz,Leskovec:2012gb}.

To evaluate $\wt F^{(i)}$, we follow the same method as described in 
Appendix B of Ref.~\cite{\dwave}, except here we have $\ell=0,1$, whereas
that work considered $\ell=0,2$.
In addition, here we extend the implementation to nonzero $\bm P$.
The sum and integral are evaluated for a value of $\alpha$ that is sufficiently small that,
based on numerical experiments, the residual dependence on $\alpha$ lies below the
desired precision for solving the quantization condition. We typically find that $\alpha=0.5$ is
adequate.

The sum in \Cref{eq:Ftb} is evaluated as written, except that 
in practice, we must introduce a cutoff, $|\bm n| < n_{\rm max}$. 
We use the same cutoff as described in Appendix B.2 of
Ref.~\cite{\dwave}, taking the $\ell'=\ell=0$ value for all choices of angular momenta.
Note that for the determination of the cutoff, the value of $|\bm n_{pP}|$ is not relevant as it
lies well below $n_{\rm max}$.

The integral in \Cref{eq:Ftb} can be evaluated analytically, as discussed in Appendix B.1
of Ref.~\cite{\dwave}. 
Focusing on the core integral, we have
\begin{align}
I^F_{\ell' m';\ell m} &= 
{\rm PV} \! \int d^3n_a \frac{e^{\alpha(x^2-r^2)}}{x^2-r^2} 
4\pi \cY_{\ell' m'}(\bm r) \cY_{\ell m}(\bm r)
\\
&=\gamma \; {\rm PV} \! \int d^3r \frac{e^{\alpha(x^2-r^2)}}{x^2-r^2} 
4\pi \cY_{\ell' m'}(\bm r) \cY_{\ell m}(\bm r)
\\
&= \delta_{\ell' \ell} \delta_{m' m} I^F_\ell\,,
\end{align}
where in the second line we have changed variables from $\bm n_a$ to $\bm r$.
The results for $I^F_0$ and $I^F_2$ are given in Eqs.~(B.4) and (B.5) of Ref.~\cite{\dwave}.
Here we also need the result for $\ell=1$,
\begin{equation}
I^F_1 = 4 \pi \gamma \left[ - \sqrt{\frac{\pi}{\alpha^3}} \frac{1+ 2 \alpha x^2}4 e^{\alpha x^2} +
\frac{\pi x^3}2 {\rm Erfi}\!\left(\sqrt{\alpha x^2}\right)
\right]
\,.
\end{equation}

\subsection{Projections}
\label{app:GT}
%

In this appendix, we detail how the block-diagonal structure of the irrep projection matrices $\{\wh P_I\}$ defined in \Cref{sec:projections} governs the irrep decomposition of the eigenvalues of the quantization condition matrix $\wh F_3^{-1} + \wh\cK_{\df,3}$, generalizing the discussion of Section 3.1 in Ref.~\cite{\dwave} to include the $\ell=1$ partial wave and all frame types.

Recall that each $\wh P_I$ is diagonal in its flavor indices $i\in\{1,2\}$ and partial waves%
\footnote{While in the main text we only discuss an implementation including $s$ and $p$ waves (consistent with a first-order threshold expansion in $\Kdf$), here we also include $d$ waves to give the moving-frame generalizations of Table 1 of Ref.~\cite{\dwave}.}
$\ell\in\{0,1,2\}$, as well as block diagonal in its spectator-momentum indices $\bm k$, with each block corresponding to a different finite-volume orbit 
 ${o_{\bm k}\equiv\{R\bm k | R\in {\rm LG}(\bm P)\}}$.
We label the innermost blocks $P^{(i)}_{I,o(\ell)}$, choosing the organization of the matrix layers to be
\begin{align}
	\wh P_I &= \text{diag} \left( P_I^{(1)} ,\; P_I^{(2)} \right) \,, \nonumber
	\\
	P_I^{(i)} &= \text{diag} \left( P^{(i)}_{I,o_1} ,\; P^{(i)}_{I,o_2} ,\; \ldots  \right) \,,
	\\ 
	P^{(i)}_{I,o} &= \text{diag} \left( P^{(i)}_{I,o(0)} ,\; P^{(i)}_{I,o(1)} ,\; P^{(i)}_{I,o(2)} \right) \,, \nonumber
\end{align}
where
\begin{align}
	\left[P^{(i)}_{I,o(\ell)}\right]_{pm';km} &= \frac{d_I}{[\text{LG}(\bm P)]} \sum_{R\in \text{LG}(\bm P)} \delta_{\bm p, R\bm k} \; \chi_I(R) \Pi(R) \cD^{(\ell)}_{m'm}(R) \label{eq:Pblock}
\end{align}
has indices corresponding to $\bm p,\bm k\in o$ and $m',m \in\{-\ell,\ldots,\ell\}$.
The orbits $o$ which contribute to a given $P^{(i)}_I$ are said to be ``active," and are precisely those for which the spectator cutoff function $H^{(i)}(\bm k\in o)$ is nonzero.
In this regard, we note that it follows 
from \Cref{eq:Hinew,eq:newzi} that $H^{(i)}(\bm k)=H^{(i)}(R\bm k)$ 
for all $R\in\text{LG}(\bm P)$, implying that the cutoff function
takes the same value for all momenta in an orbit.
Since each orbit $o$ is defined with respect to a particular little group $\text{LG}(\bm P)$, 
different values $\bm P$ yield different types of orbits.
We list the orbit types that arise for all classes of frame in \Crefrange{tab:o1}{tab:o6}.
We label frames using the notation $\bm P = \bm d_{\rm ref} (2\pi/L)$,
where $\bm d_{\rm ref}$ is a vector of integers.
No table is shown for $\bm d_{\rm ref}=(n_1,n_2,n_3)$, with $n_1$, $n_2$ and $n_3$ distinct,
nonzero integers,
since the corresponding little group is trivial, and all orbits are one-dimensional.
\begin{table}[tb]
\centering
\scalebox{0.91}{
\begin{tabular}{c|c|c}
\text{orbit}&\text{size}   & \text{elements}  \\
\hline
$o_{000}$& 1 	& $(0,0,0)$ \\
$o_{00a}$& 6 	& $(\pm a,0,0),\, (0,\pm a,0),\, (0,0,\pm a)$ \\
$o_{aa0}$&  12 	& $(\pm a,\pm a,0),\, (\pm a,0,\pm a),\, (0,\pm a,\pm a)$ \\
$o_{aaa}$&8 	& $(\pm a,\pm a,\pm a)$ \\
$o_{ab0}$&24 	& $(\pm a,\pm b,0),\, (\pm a,0,\pm b),\, (0,\pm a,\pm b),\, (\pm b,\pm a,0),\, (\pm b,0,\pm a),\, (0,\pm b,\pm a)$ \\
$o_{aab}$ &24 	& $(\pm a,\pm a,\pm b),\, (\pm a,\pm b,\pm a),\, (\pm b,\pm a,\pm a)$ \\
{$o_{abc}$}&48 	& 
$(\pm a,\pm b,\pm c),\, (\pm a,\pm c,\pm b),\, (\pm b,\pm a,\pm c),\,
(\pm b,\pm c,\pm a),\, (\pm c,\pm a,\pm b),\, (\pm c,\pm b,\pm a)$ \\
\end{tabular}
}
\caption{Explicit compositions of all types of orbit for the rest frame, $\bm d_{\rm ref}=(0,0,0)$.
Listed are the forms of the spectator-momentum vectors $\bm n_k= \bm k (L/2\pi)$.
The integers $a,b,c$ satisfy $0<a<b<c$ but are otherwise arbitrary, 
and all $\pm$ signs are independent of each other.
\label{tab:o1}}
\end{table}

\begin{table}[tb]
\centering
\begin{tabular}{c|c|c}
\text{orbit} & \text{size}   & \text{elements}  \\
\hline
$o_{00z}$ & 1 	& $(0,0,z)$ \\
$o_{a0z}$ & 4 	& $(\pm a,0,z),\, (0,\pm a,z)$ \\
$o_{aaz}$ & 4 	& $(\pm a,\pm a,z)$ \\
$o_{abz}$ & 8 	& $(\pm a,\pm b,z),\, (\pm b,\pm a,z)$
\end{tabular}
\caption{Orbit types for the frame with $\bm d_{\rm ref}=(0,0,n)$, where $n\ne 0$.
Notation as in \Cref{tab:o1}, with
$0<a<b$, and $z$ an unconstrained integer (which may be positive, negative, or zero).
\label{tab:o2}}
\end{table}

\begin{table}[tb]
\centering
\begin{tabular}{c|c|c}
\text{orbit} & \text{size}   & \text{elements}  \\
\hline
$o_{xx0}$ & 1 		& $(x,x,0)$ \\
$o_{xxa}$ & 2 		& $(x,x,\pm a)$ \\
$o_{xy0}$ & 2 		& $(x,y,0),\, (y,x,0)$ \\
$o_{xya}$ & 4 		& $(x,y,\pm a),\, (y,x,\pm a)$
\end{tabular}
\caption{Orbit types for the frame with $\bm d_{\rm ref}=(n,n,0)$, where $n\ne 0$.
Notation as in \Cref{tab:o1}, with integers $a,x,y$ satisfying  $a>0$ and $y > x$.
\label{tab:o3}}
\end{table}

\begin{table}[tb]
\centering
\begin{tabular}{c|c|c}
\text{orbit} & \text{size}   & \text{elements}  \\
\hline
$o_{xxx}$ & 1 		& $(x,x,x)$ \\
$o_{xxy}$ & 3 		& $(x,x,y),\, (x,y,x),\, (y,x,x)$ \\
$o_{xyz}$ & 6 		& $(x,y,z),\, (x,z,y),\, (y,x,z),\, (y,z,x),\, (z,x,y),\, (z,y,x)$ \\
\end{tabular}
\caption{Orbit types for the frame with $\bm d_{\rm ref}=(n,n,n)$, where $n\ne 0$.
Notation as in \Cref{tab:o1}, with $x$, $y$, and $z$ being distinct integers.
For $\sigma_{xyz}$, we take $|x|\leq|y|\leq|z|$ (supplemented by the condition that if, e.g.~$x=-y$, then
we choose the ordering such that $x<0<y$),
unless exactly one of them is zero, in which case we take $0=z<|x|\leq|y|$ (again with $x<0<y$ if $x=-y$).
\label{tab:o4}}
\end{table}

\begin{table}[tb]
\centering
\begin{tabular}{c|c|c}
\text{orbit} & \text{size}   & \text{elements}  \\
\hline
$o_{xy0}$ & 1 		& $(x,y,0)$ \\
$o_{xya}$ & 2 		& $(x,y,\pm a)$ \\
\end{tabular}
\caption{Orbit types for the frame with
$\bm d_{\rm ref}=(n_1, n_2,0)$, where $n_1$ and $n_2$ are distinct, nonzero integers.
Here $a>0$, while $x$ and $y$ are unconstrained integers that can be equal.
\label{tab:o5}}
\end{table}

\begin{table}[tb]
\centering
\begin{tabular}{c|c|c}
\text{orbit} & \text{size}   & \text{elements}  \\
\hline
$o_{xxz}$ & 1 & $(x,x,z)$ \\
$o_{xyz}$ & 2 & $(x,y,z),\, (y,x,z)$
\end{tabular}
\caption{Orbit types for the frame with $\bm d_{\rm ref}=(n_1,n_1,n_2)$, 
where $n_1$ and $n_2$ are distinct, nonzero integers.
Here $x\neq y$, while $z$ is unconstrained.
\label{tab:o6}}
\end{table}

Since $\wh P_I$ a projection matrix, each of its eigenvalues is either zero or one.
The dimension of the projected irrep subspace is given by the number of unit eigenvalues, i.e.
\begin{align}
	d(\wh P_I) = \text{Tr}(\wh P_I)
	= \sum_{i=1}^2 \sum_o \sum_{\ell=0}^2 d(P^{(i)}_{I,o(\ell)}) \,.
\end{align}
The sub-block dimensions $d(P^{(i)}_{I,o(\ell)}) = \text{Tr}(P^{(i)}_{I,o(\ell)})$ can be directly computed from \Cref{eq:Pblock}, and are given in Tables~\ref{tab:d1}--\ref{tab:d6} for each orbit type of each little group $\text{LG}(\bm P)$.
The results for $\ell=0,2$ in Table~\ref{tab:d1} are from Ref.~\cite{\dwave}; those for $\ell=1$ are new.
The other tables (for the moving frames) are also new to this work.

\begin{table}[tb]
\centering
\scalebox{0.93}{
\begin{tabular}{c|ccccccc}
 &\multicolumn{7}{c}{orbit types}\\ 
\text{irrep} & 
$(000)_1$    &  $(00a)_6$   & $(aa0)_{12}$ & $(aaa)_8$ & $(ab0)_{24}$ & $(aab)_{24}$ & $(abc)_{48}$ \\
 \hline
$ A_1^+[1]$ & (0,0,0) & (0,0,0) & (0,0,1)  & (0,0,0) & (0,1,2)  & (0,1,2) & (1,3,5)     \\
$ A_2^+[1]$ & (0,0,0) & (0,0,1) & (0,1,1)  & (1,1,1) & (0,1,2)  & (1,2,3) & (1,3,5)     \\
$ E^+[2]$ & (0,0,0) & (0,0,2) & (0,2,4)  & (0,2,4) & (0,4,8)  & (2,6,10) & (4,12,20)   \\
$ T_1^+[3]$ & (0,3,0) & (3,6,6) & (3,9,12) & (3,6,9) & (6,15,24) & (6,15,24) & (9,27,45)   \\
$ T_2^+[3]$ & (0,0,0) & (0,3,6) & (3,6,12) & (0,3,6) & (6,15,24) & (3,12,21) & (9,27,45)    \\
$A_1^-[1]$ & (1,0,0) & (1,1,1) & (1,1,2)  & (1,1,1) & (1,2,3)  & (1,2,3)   & (1,3,5)   \\
$ A_2^-[1]$ & (0,0,0) & (0,0,1) & (0,1,1)  & (0,0,0) & (1,2,3)  & (0,1,2)   & (1,3,5)   \\
$ E^-[2]$ & (0,0,2) & (2,2,4) & (2,4,6)  & (0,2,4) & (4,8,12) & (2,6,10)  & (4,12,20)  \\
$ T_1^-[3]$ & (0,0,0) & (0,3,3) & (0,6,9)  & (0,3,6) & (3,12,21) & (3,12,21)  & (9,27,45)  \\
$ T_2^-[3]$ & (0,0,3) & (0,3,6) & (3,6,12) & (3,6,9) & (3,12,21) & (6,15,24)  & (9,27,45)  \\
\hline
total & (1,3,5) & (6,18,30) & (12,36,60) & (8,24,40) & (24,72,120) & (24,72,120) & (48,144,240) \\ 
\end{tabular}
}
\caption{Dimension of irrep projection sub-blocks for each orbit type
and angular momentum, $(d(P_{I,o(0))},  d(P_{I,o(1)}), d(P_{I,o(2)}))$, for the frame
with $\bm d_{\rm ref}=(0,0,0)$.
Each row corresponds to an irrep of the cubic group $O_h$, 
whose dimension is listed in square parentheses.
Entries assume particles are pseudoscalars; for scalars irreps should be interchanged
with their parity counterparts.
See \Cref{tab:o1} for orbit-type compositions.
The bottom row gives the sum of the rows above,
 which equals $d_{\rm orbit} (2\ell+1)$, 
 where $d_{\rm orbit}$ is the number of elements in the orbit,
 which is given as a subscript for each orbit type.
\label{tab:d1}}
\end{table}

\begin{table}[tb]
\centering
\begin{tabular}{c|cccc}
& \multicolumn{4}{c}{orbit types}\\
\text{irrep}   & $(00z)_1$   & $(a0z)_4$ & $(aaz)_4$ & $(abz)_8$  \\
\hline
$A_1[1]$ & (0,\,0,\,0) & (0,\,1,\,2) & (0,\,1,\,2) & (1,\,3,\,5) \\
$A_2[1]$ & (1,\,1,\,1) & (1,\,2,\,3) & (1,\,2,\,3) & (1,\,3,\,5) \\ 
$B_1[1]$ & (0,\,0,\,1) & (0,\,1,\,2) & (1,\,2,\,3) & (1,\,3,\,5) \\ 
$B_2[1]$ & (0,\,0,\,1) & (1,\,2,\,3) & (0,\,1,\,2) & (1,\,3,\,5) \\
$E[2]$  & (0,\,2,\,2) & (2,\,6,\,10)& (2,\,6,\,10)& (4,\,12,\,20) \\
\hline
total &  (1,\,3,\,5) & (4,\,12,\,20)& (4,\,12,\,20)& (8,\,24,\,40)
\end{tabular}
\caption{As for Table~\ref{tab:d1} but for frames with $\bm d_{\rm ref}=(0,0,n)$.
See \Cref{tab:o2} for orbit-type compositions.
\label{tab:d2}}
\end{table}

\begin{table}[tb]
\centering
\begin{tabular}{c|cccc}
 &\multicolumn{4}{c}{orbit types}\\
\text{irrep}   & $(xx0)_1$    &  $(xxa)_2$   & $(xy0)_2$ & $(xya)_4$ \\
\hline
$A_1[1]$ & (0,\,0,\,1) & (0,\,1,\,2) & (0,\,1,\,2) & (1,\,3,\,5) \\
$A_2[1]$ & (1,\,1,\,2) & (1,\,2,\,3) & (1,\,2,\,3) & (1,\,3,\,5) \\ 
$B_1[1]$ & (0,\,1,\,1) & (0,\,1,\,2) & (1,\,2,\,3) & (1,\,3,\,5) \\ 
$B_2[1]$ & (0,\,1,\,1) & (1,\,2,\,3) & (0,\,1,\,2) & (1,\,3,\,5) \\
\hline
total &    (1,\,3,\,5) & (2,\,6,\,10)& (2,\,6,\,10)& (4,\,12,\,20)
\end{tabular}
\caption{As for Table~\ref{tab:d1} but for frames with $\bm d_{\rm ref}=(n,n,0)$.
See \Cref{tab:o3} for orbit-type compositions.
\label{tab:d3}}
\end{table}

\begin{table}[tb]
\centering
\begin{tabular}{c|ccc}
& \multicolumn{3}{c}{orbit types}\\
\text{irrep}    & $(xxx)_1$    &  $(xxy)_3$   & $(xyz)_6$ \\
\hline
$A_1[1]$ & (0,\,0,\,0) & (0,\,1,\,2) & (1,\,3,\,5) \\
$A_2[1]$ & (1,\,1,\,1) & (1,\,2,\,3) & (1,\,3,\,5) \\
$E[2]$ & (0,\,2,\,4) & (2,\,6,\,10)& (4,\,12,\,20)  \\ 
\hline
total & (1,\,3,\,5) & (3,\,9,\,15)& (6,\,18,\,30)
\end{tabular}
\caption{As for Table~\ref{tab:d1} but for frames with $\bm d_{\rm ref}=(n,n,n)$.
See \Cref{tab:o4} for orbit-type compositions.
\label{tab:d4}}
\end{table}

\begin{table}[tb]
\centering
\begin{tabular}{c|cc}
&  \multicolumn{2}{c}{orbit types}\\
\text{irrep}   & $(xy0)_1$    &  $(xya)_2$  \\
\hline
$A_1[1]$ & (0,\,1,\,2) & (1,\,3,\,5) \\
$A_2[1]$ & (1,\,2,\,3) & (1,\,3,\,5) \\
\hline
total &   (1,\,3,\,5) & (2,\,6,\,10)
\end{tabular}
\caption{As for Table~\ref{tab:d1} but for frames with $\bm d_{\rm ref}=(n_1,n_2,0)$.
See \Cref{tab:o5} for orbit-type compositions.
\label{tab:d5}}
\end{table}

\begin{table}[tb]
\centering
\begin{tabular}{c|cc}
 &\multicolumn{2}{c}{orbit types}\\
\text{irrep}    & $(xxz)_1$    &  $(xyz)_2$  \\
\hline
$A_1[1]$ & (0,\,1,\,2) & (1,\,3,\,5) \\
$A_2[1]$ & (1,\,2,\,3) & (1,\,3,\,5) \\
\hline
total &   (1,\,3,\,5) & (2,\,6,\,10)
\end{tabular}
\caption{As for Table~\ref{tab:d1} but for $\bm P=(n_1,n_1,n_2)$.
See \Cref{tab:o6} for explicit orbit-type compositions.
\label{tab:d6}}
\end{table}

Although the quantization condition matrix $\wh F_3^{-1} + \wh\cK_{\df,3}$ does not have the same block-diagonal structure as the projectors, the precise irrep decomposition of its eigenvalues can still be determined from the appropriate table.
This is best illustrated with a concrete example.
Consider a finite-volume 2+1 system of pseudoscalars with $m_1L=4$, $m_2L=6$,
in a moving frame with $\bm d_{\rm ref} = (0,0,1)$, 
with total CMF energy ${E^*=(2m_1 + m_2) + m_1/2 = 4m_1}$, 
and including $s$- and $p$-wave dimers when the spectator has flavor 1 
but only $s$-wave dimers when the spectator has flavor 2.
The flavor-1 and flavor-2 spectator momenta that are active 
($H^{(i)}(\bm p_i)>0$) for this system are 
\begin{equation}
\begin{split}
	\bm n_{p_1} &\in\{(0,0,0), (0,0,1)\} = o_{000} \cup o_{001}
	\\
	\bm n_{p_2} &\in\{(0,0,0), (0,0,1),(1,0,1),(0,1,1),(-1,0,1),(0,-1,1)\} = o_{000} \cup o_{001} \cup o_{101} \,,
	\end{split}
\end{equation}
where the orbit notation is that of \Cref{tab:o2}; $o_{000}$ and $o_{001}$ are of type $o_{00z}$, while $o_{101}$ is of type $o_{a0z}$.
Each value of $\bm n_{p_1}$ brings $1+3=4$ eigenvalues ($s$- and $p$-waves), while each value of $\bm n_{p_2}$ only brings a single ($s$-wave) eigenvalue, giving a total $2\cdot 4 + 6 = 14$ eigenvalues of $\wh F_3^{-1} + \wh\cK_{\df,3}$, i.e.~it is a $14\times14$ matrix.

From the $(00z)_1$ and $(a0z)_4$ columns of \Cref{tab:d2}, we can compute the total number of eigenvalues of $\wh F_3^{-1} + \wh\cK_{\df,3}$ that lie in the $A_2$ irrep:
\begin{align}
	d(\wh P_{A_2}) &= 2\big[\underbrace{d(P_{A_2,00z(0)})}_1 
	+ \underbrace{d(P_{A_2,00z(1)})}_1 \big] + 
	\big[ 2\underbrace{d(P_{A_2,00z(0)})}_1 
	+ \underbrace{d(P_{A_2,a0z(0)})}_1 \big] = 7 \,.
\end{align}
Similarly, we find that there is 1 eigenvalue in $B_2$ and 6 in $E$ (i.e.~three doublets), giving the correct total of 14 eigenvalues of $\wh F_3^{-1} + \wh\cK_{\df,3}$.

\section{Threshold expansion in the nondegenerate case}
\label{app:threshold}

In this appendix we derive the threshold expansion for three nonidentical scalars
that have, in general, different masses and interactions.
The threshold expansion refers to the $1/L$ expansion of the ground (or threshold) 
state of few-particle systems. 
Here we work to next-to-next-to-leading order (NNLO), 
determining the first three terms in the expansion.
We consider only the case of vanishing total momentum, $\bm P=0$.

The motivation for this derivation is twofold.
First, it provides a way of checking our
numerical implementation of the quantization condition for nondegenerate systems.
Second, it provides a check of the formalism itself, since, based on previous
results for identical particles and the 2+1 system, we have a very clear expectation for
the form of the threshold expansion for the nondegenerate case.

The threshold expansion has been derived previously for systems of identical particles~\cite{Luscher:1986n2,Beane:2007es,Hansen:2016fzj,Pang:2019dfe,Romero-Lopez:2020rdq}, 
for three-pion systems with $I=1$~\cite{Muller:2020vtt},
and for the $n\pi^+ + m K^+$ system~\cite{Smigielski:2008pa}. 
The fully nondegenerate three-particle system has not been previously considered.
The methods used have varied, and include nonrelativistic quantum mechanics (NRQM)
as well as starting from a relativistic quantization condition.
It is known from the identical-particle case that, up to NNLO, the results from all
approaches agree.

The generic $n\pi^+ + m K^+$ result of Ref.~\cite{Smigielski:2008pa} includes the
$\pi^+\pi^+K^+$ and $\pi^+K^+K^+$ systems as special cases.
Thus we can use these results to check our numerical implementation of the
$2+1$ quantization condition.
We can, however, also use them to provide a check of the formalism itself.
Since the results of Ref.~\cite{Smigielski:2008pa} were obtained using NRQM,
it is a nontrivial check of the formalism described in \Cref{sec:formalism}
that it reproduces the threshold expansion.
In particular, it checks the various symmetry factors in the expressions.
Thus we have also derived the threshold expansion in the 2+1 case, and we comment
briefly on this at the end of this section.

We denote the masses of the three particles by $m_1$, $m_2$, and $m_3$.
We know from Ref.~\cite{Beane:2003yx} that the leading order energy shift for
the {\em two-particle} ground state composed of flavors $j$ and $k$ is
\begin{equation}
\Delta E^{(i)}_2 = E_2^{(i)} - m_j - m_k = \frac{2 \pi a_0^{(i)}}{\mu_i L^3} + \cO(L^{-4})\,,
\label{eq:deltaE2}
\end{equation}
where $i$, $j$, and $k$ are ordered cyclically, and
\begin{equation}
\mu_i = \frac{m_j m_k}{m_j + m_k}
\end{equation}
is the reduced mass of the $j$, $k$ pair.
Thus a reasonable expectation for the leading-order shift for the three-particle threshold state is
\begin{equation}
\Delta E = E - m_1 - m_2 - m_3 = \sum_{i=1}^3 \frac{2 \pi a_0^{(i)}}{\mu_i L^3}\,.
\label{eq:LOthreshold}
\end{equation}
Indeed, we will see that this result is reproduced by expanding the quantization condition.

In the remainder of this section, we determine the constants $c_3$, $c_4$, and $c_5$
in the general threshold expansion,
\begin{equation}
\Delta E = \frac{c_3}{L^3} + \frac{c_4}{L^4} + \frac{c_5}{L^5} + \cO(L^{-6}) \,.
\end{equation}
%
%
The previous derivation of the threshold expansion using the RFT form of the quantization condition 
was carried out in Ref.~\cite{\HSTH} for identical particles. The expansion was obtained to NNNLO,
and most of the effort in the analysis was devoted to obtaining the $c_6/L^6$ term,
which is the lowest order at which $\Kdf$ enters the expansion.
Here we work only to NNLO, and it turns out that a much simpler approach suffices.
We have, however, checked our results by also following the method of derivation of
Ref.~\cite{\HSTH}.

The starting point of the derivation is the symmetric form of the quantization condition
for nondegenerate particles obtained in Ref.~\cite{Blanton:2020gmf}. 
This has the same form as \Cref{eq:QC3}, with $\wh F_3$ given by \Cref{eq:F3hat},
except that the flavor space is now three-dimensional.
Equations (\ref{eq:Fhat})--(\ref{eq:K2Lhat}) are replaced by
\begin{align}
\wh F &= {\rm diag}\left( \wt F^{(1)}, \wt F^{(2)}, \wt F^{(3)}\right)\,,
\label{eq:FhatND}
\\
\wh G &= 
\begin{pmatrix}0 & \wt G^{(12)} P_L & P_L \wt G^{(13)}\\
P_L \wt G^{(21)} & 0 & \wt G^{(23)} P_L\\
\wt G^{(31)} P_L & P_L \wt G^{(32)}  & 0 \\
\end{pmatrix}\,,
\label{eq:GhatND}
\\
\wh{\overline \cK}_{2,L} &= {\rm diag} \left(
{\overline \cK}_{2,L}^{(1)}, {\overline \cK}_{2,L}^{(2)}, {\overline \cK}_{2,L}^{(3)}
\right)\,,
\label{eq:K2LhatND}
\end{align}
where the three-flavor quantities $\wt F^{(j)}$, $\wt G^{(ij)}$, and ${\overline \cK}_{2,L}^{(j)}$
are defined by the obvious generalizations of
the expressions given in \Cref{sec:formalism}; explicit expressions are collected in 
Appendix A of Ref.~\cite{\BStwoplusone}.

Since $\Kdf$ does not enter the threshold expansion at NNLO, we can set it to zero.
Then the quantization condition reduces to $\det \wh F_3^{-1} = 0$.
Given the form of $\wh F_3$, \Cref{eq:F3hat}, the solutions to this equation 
are those that occur at free energies (which is where $\wh F$ has poles),
and those depending on the two-particle interactions that occur when
\begin{equation}
\det \widehat H  = 0, \quad \text{ with } \quad 
\widehat H =  \widehat{\overline{\cK}}_{2,L} + \widehat{F} + \widehat{G}\,.
\label{eq:simpleQC3}
\end{equation}
As discussed in \Cref{sec:formalism}, the solutions at free energies are spurious.
Thus we can use the simpler form of the quantization condition, \Cref{eq:simpleQC3}, for
our analysis.

Another result that we can take over from Ref.~\cite{\HSTH} is that, up to NNLO,
only the $s$-wave matrix elements of $\widehat H$ are relevant for the energy shift. 
This is because higher waves are suppressed by barrier factors that bring in
additional powers of momenta, and these momenta scale as $1/L$.
Thus the only nontrivial matrix index aside from the flavor index is that of the spectator
momentum, so we use the reduced notation exemplified by $[\wh H]_{p k}$
and $\wt F^{(i)}_{p k}$.

\subsection{Expansions of kinematic functions and $\wh{\overline{\cK}}_{2,L}$}

We will need the large volume expansions of the quantities appearing in $\wh H$.
Starting with $\wt F^{(i)}$, and recalling that $\bm P=0$,
we have from \Cref{eq:Ftb} that
\begin{equation}
\widetilde F^{(i)}_{00} = \frac{1}{16 \pi^2 L^4 m_i (E - m_i) x^2 } \left[ 1 + \left( \sum_{\bf n \neq 0} - \text{PV}\int d^3 n \right) \frac{x^2 e^{\alpha(x^2-n^2)}}{x^2 - n^2} \right], \label{eq:Fexpansion}
\end{equation}
where $x = q_{2,p}^{*(i)} L /(2 \pi) $ with $\bm p=0$. Note that the dependence on the masses of the
scattering pair, $m_j$ and $m_k$, enters only through $x^2$.
The UV is regulated by $\alpha>0$~\cite{\KSS}, and one sends $\alpha\to 0^+$ in the end.
From Ref.~\cite{\HSTH} we have the result
\begin{equation}
\lim_{\alpha\to 0^+}\left(\sum_{\bm n} - {\rm PV}\!\!\int d^3n\right) 
\frac{x^2 e^{\alpha(x^2-n^2)}}{x^2-n^2}= - x^2 \cI - x^4 \cJ + \cO(x^6)\,,
\end{equation}
where $\cI$ and $\cJ$ are known numerical constants. 
Using the kinematic results
\begin{align}
\begin{split}
E - m_i &= m_j + m_k + \Delta E = m_j + m_k + O(L^{-3}), \\ 
q_{2,p=0}^{*(i)2} &= 2 \mu_i \Delta E \left[ 1 + O(L^{-3}) \right],
\end{split}
\end{align}
we then find
\begin{equation}
\widetilde F^{(i)}_{00} = \frac{1}{L^3} \frac{1}{8 m_1 m_2 m_3}
 \left[ \frac{1}{L^3 \Delta E} - \frac{\mu_i}{2 \pi^2 L} \mathcal I 
 - \left( \frac{\mu_i}{2 \pi^2 L} \right)^2 L^3 \Delta E  \mathcal J  
 + \hdots \right]\,.
 \label{eq:FwithJ}
\end{equation}
Since $L^3\Delta E \sim L^0$,
the leading term in $\wt F^{(i)}_{00}$ is of $\cO(1/L^3)$, 
with corrections suppressed by powers of $1/L$.
We also need the result that $\wt F^{(i)}_{pp}$ for $\bm p\ne 0$
scales as $1/L^4$ if $|\bm p|$ is treated as of $\cO(1/L)$~\cite{\HSTH}.

The expansion of $\widetilde G^{(ij)}$ can be obtained from \Cref{eq:Gt}.
We find
\begin{equation}
\widetilde G_{00}^{(ij)} = \frac{1}{4 m_i m_j L^6} \frac{1}{(E-m_i -m_j)^2 - m_k^2} = \frac{1}{L^3}\frac{1}{8 m_i m_j m_k} \frac{1}{\Delta E L^3} + O(L^{-6}),
\label{eq:G00}
\end{equation} 
while the off-diagonal elements are suppressed by an additional power of $1/L$,
\begin{equation}
\wt G_{0p}^{(ij)}  = \wt G_{p0}^{(ji)} 
= -\frac{1}{L^3} \frac{1}{8 m_1 m_2 m_3} \frac{\mu_{i}}{2 \pi^2 L} \frac{1}{n^2} + O(L^{-4})\,,
\end{equation}
where $\bm n\equiv \bm p L/(2\pi)$ is treated as of $\cO(L^0)$.
Note that the flavor label of the reduced mass matches that of the spectator flavor with zero momentum.
We also need the result that $\wt G_{pp}^{(ij)}$ scales as $1/L^4$~\cite{\HSTH}.

Finally, the expansion of $1/\overline{\cK}_{2,L}^{(i)}$ can be obtained from
\Cref{eq:K2Li,eq:K2,eq:EREs}. 
Using the results that,
for $\bm p= \bm n (2\pi/L)$ with $|\bm n| \sim L^0$,
 we have $q_{2,p}^{*(i)2} = O(L^{-2})$, $\omega_p^{(i)}= m_i + O(L^{-2})$
and $\sqrt{\sigma_i}=m_j+m_k + O(L^{-2})$, we find
\begin{equation}
\frac{1}{[\overline{\cK}_{2,L}^{(i)}]_{pp}} = -
 \frac{1}{L^3}\frac{1}{8 m_i m_j m_k} \frac{\mu_i}{2 \pi a_0^{(i)}}+ O(L^{-5}). 
 \label{eq:K2expansion}
\end{equation}

\subsection{Derivation to $O(L^{-4})$}

To obtain the energy shift at NLO, we can restrict to zero spectator momentum, 
and solve $\det \wh H_{00} = 0$ to determine $\Delta E$. 
Qualitatively, this restriction is justified because contributions with nonzero spectator momenta
are suppressed by powers of $1/L$.
It turns out, as will be seen from the full NNLO calculation, that the $c_3$ and $c_4$ contributions
are obtained correctly in this way. The full matrix is given by
\begin{equation}
\wh H_{00} =\begin{pmatrix}
\widetilde F^{(1)}_{00} +{1}/{[\overline \cK_{2,L}^{(1)}]_{00}} 
& \widetilde G_{00}^{(12)}& \widetilde G_{00}^{(13)} 
\\
\widetilde G_{00}^{(21)} & \widetilde F^{(2)}_{00} +{1}/{[\overline \cK_{2,L}^{(2)}]_{00}}  
& \widetilde G_{00}^{(23)} 
\\
\widetilde G_{00}^{(31)}& \widetilde G_{00}^{(32)} 
& \widetilde F^{(3)}_{00} +{1}/{[\overline \cK_{2,L}^{(3)}]_{00}}  
\end{pmatrix}\,,
\label{eq:H00}
\end{equation}
where we keep only the leading and $1/L$ suppressed terms from 
\Cref{eq:FwithJ,eq:G00,eq:K2expansion}.
After some algebra, we find that requiring the determinant to vanish leads to
\begin{equation}
 L^3 \Delta E = \sum_{i=1}^3 \frac{2 \pi a_0^{(i)}}{\mu_i} 
 \left(1 - \frac{a_0^{(i)}}{\pi L} \mathcal I \right).
\end{equation}
We see that the energy shift is given by pairwise interactions at NLO.

\subsection{Derivation to $O(L^{-5})$}

To extend the result to NNLO, we must include all entries in $\wh H$.
It is convenient to 
divide the matrices in sub-blocks according to whether the spectator momentum is zero or nonzero:
\begin{equation}
\wh H = \begin{pmatrix}
\wh H_{00} & \wh H_{0p} \\ \wh H_{p0} & \wh H_{pp} 
\end{pmatrix}\,.
\end{equation}
Here $\wh H_{00}$ is as given in \Cref{eq:H00}, except now we include $1/L^2$ suppressed
terms (which enter only in $\wt F^{(i)}_{00}$),
while
\begin{equation}
\wh H_{0p} =
\begin{pmatrix}
0 & \wt G_{0p}^{(12)}& \wt G_{0p}^{(13)} \\
\wt G_{0p}^{(21)}& 0 & \wt G_{0p}^{(23)} \\
\wt G_{0p}^{(31)}& \wt G_{0p}^{(32)}& 0 
\end{pmatrix}\,,
\label{eq:H0k}
\end{equation}
with $\wh H_{p0}$ given by interchanging the momentum indices,
and
\begin{equation}
 \wh H_{pp} = \begin{pmatrix} 1/[\overline \cK_{2,L}^{(1)} ]_{pp}&0&0\\
  0&1/[\overline \cK_{2,L}^{(2)}]_{pp}& 0\\
  0&0& 1/[\overline \cK_{2,L}^{(3)}]_{pp} \end{pmatrix} + O(L^{-4})\,.
\end{equation}
We stress that the abbreviation $p$ implies that all finite-volume, nonzero values of $\bm p$
are to be included.
Thus, for example,
 $\wt G^{(ij)}_{0p}$ is a rectangular matrix (in fact, a row vector) in momentum space,
 while $1/[\overline \cK_{2,L}^{(1)} ]_{pp}$ is a square matrix (with only diagonal entries).\footnote{%
 The cutoff functions $H^{(i)}(\bm p)$ do not appear at any order in the $1/L$ expansions,
 and so, formally, there is no cutoff on $\bm p$, and these matrices are infinite dimensional.
 This is not a concern, however, because the sum over $\bm p$ that appears below is
 convergent.}
 We note that the $\wh F$ and $\wh G$ contributions to $\wh H_{pp}$ are part of the
 $O(L^{-4})$ term, and are not needed at the order we work.

We can now use the following result for the determinant of a block matrix:
\begin{equation}
\det \wh H =  \det(\wh H_{00}  - \wh H_{0p}  \wh H_{pp}^{-1}\wh  H_{p0} ) \det \wh H_{pp}\,,
\end{equation}
valid as long as $\wh H_{pp}$ is invertible, as is the case here.
The second term on the right-hand side involves only infinite-volume two-particle K matrices
(aside from overall factors of $L$) and thus cannot lead to a finite-volume quantization condition.
The latter is instead obtained from the first term, leading to
\begin{equation}
 \det(\wh H_{00}  - \wh H_{0p}  \wh H_{pp}^{-1}\wh  H_{p0} )  = 0\,.
 \label{eq:QC3NNLO}
\end{equation}
We note that the second term in this determinant is
suppressed by $1/L^2$ compared to the first, thus justifying the
use of $\det \wh H_{00}=0$ when working at NNLO.

It only remains to compute the determinant keeping all terms with a relative suppression of $1/L^2$. 
A complication is that the term with $\mathcal J$ in \Cref{eq:FwithJ} also depends on the energy shift. 
However, it is consistent to replace $L^3 \Delta E$ in the last term of \Cref{eq:Fexpansion} by the leading-order result for the energy shift, given by the $c_3$ contribution in \Cref{eq:LOthreshold}. 
After some algebra, and using $\cJ=\sum_{\bm n\ne 0} 1/(n^2)^2$, we arrive at the final result,
\begin{align}
\begin{split}
 L^3 \Delta E &= \sum_{i=1}^3 \frac{2 \pi a_0^{(i)}}{\mu_i} \left[1 - \frac{a_0^{(i)}}{\pi L} \mathcal I 
  + \left(\frac{a_0^{(i)}}{\pi L} \right)^2 (\mathcal I ^2 - \mathcal{J})  
  + \frac{a^{(j)}_0 a^{(k)}_0}{(\pi L)^2} 2\mathcal{J} \right] + O(L^{-6})\,,
\end{split}
  \label{eq:NNLOthresholdapp}
\end{align}
where $i$, $j$, and $k$ are ordered cyclically. The first three terms in the square brackets
are exactly those of the threshold expansion for two-particle energies.
The final term, however, corresponds to a three-particle contribution in which three pairs
interact in turn. 
 We observe that there are no terms of the form $a_0^{(i)} a_0^{(j)2}$, which
one might expect from a three-pair interaction. However, although these terms appear
at intermediate stages, they cancel in the final expression.

We have repeated the derivation for the 2+1 system, using the quantization condition
\Cref{eq:QC3}. We do not present the details.  The result is
\begin{align}
\begin{split}
 L^3 \Delta E &= 2 \frac{2 \pi a_0^{(1)}}{\mu_1} \left[1 - \frac{a_0^{(1)}}{\pi L} \mathcal I 
  + \left(\frac{a^{(2)}_0}{\pi L}\right)^2 (\mathcal I ^2 - \mathcal{J})  
  +   \frac{a_0^{(1)} a_0^{(2)}}{(\pi L)^2}  2\mathcal{J} \right] 
  \\
  &\quad +  \frac{2 \pi a_0^{(2)}}{\mu_2} \left[1 - \frac{a_0^{(2)}}{\pi L} \mathcal I 
  + \left(\frac{a_0^{(2)}}{\pi L} \right)^2 (\mathcal I ^2 - \mathcal{J})  
  + \left(\frac{a^{(1)}_0}{\pi L}\right)^2 2\mathcal{J} \right] 
    + O(L^{-6})\,,
\end{split}
  \label{eq:2+1TH}
\end{align}
where $\mu_1$ is the reduced mass of the $1'2$ pair, and $a_0^{(1)}$ the scattering length,
while $\mu_2$ and $a_0^{(2)}$ are the corresponding quantities for the $11'$ pair.
We note that this is exactly the result obtained
by taking the limit of \Cref{eq:NNLOthresholdapp} when two of the particles
are degenerate and have the same interactions with the third.
This agreement is nontrivial,
since in the $2+1$ system the degenerate pair are identical, whereas
in the degenerate limit of the nondegenerate case studied above the pair are distinguishable.
In particular, at a technical level, the agreement requires the cancellation of various symmetry
factors, including that in the expression for $\cK_2$, \Cref{eq:K2}.

We also note that if one takes the fully degenerate limit of \Cref{eq:NNLOthresholdapp},
with all interactions also equal,
then one reproduces the identical-particle result of Ref.~\cite{Beane:2007es}.

\section{Relating $\Kdf$ to $\cM_3$}
\label{app:KtoM}


In this appendix we generalize the definition of $\Mdf$ to the 2+1 theory, and derive the
result (\ref{eq:MtoKLO}) used in the main text.

The starting point is the result from Ref.~\cite{\BStwoplusone} for the
finite-volume version of $\cM_3$ in the 2+1 theory, which is obtained by combining
Eqs.~(86)--(88) of that work,
\begin{equation}
\cM_{3,L} = \bra{\alpha_\cS} \wh \cD_L^{(u,u)} \ket{\alpha_\cS}
+ \bra{\alpha_\cS} [1/3 - \wh \cD_{23,L}^{(u,u)} \wh F]
\wh \cK_{\df,3} \frac1{1 + \wh F_3 \wh \cK_{\df,3}}
[1/3 - \wh F \wh \cD_{23,L}^{(u,u)}]\ket{\alpha_\cS}\,.
\label{eq:MtoKL}
\end{equation}
Here
\begin{align}
 \wh \cD_L^{(u,u)} &= - \wh{\overline{\cM}}_{2,L} \wh G \wh{\overline{\cM}}_{2,L}
 \frac1{1+ \wh G \wh{\overline{\cM}}_{2,L}}\,,
 \\
  \wh \cD_{23,L}^{(u,u)} &= \wh{\overline{\cM}}_{2,L}  +  \wh \cD_L^{(u,u)}\,,
 \\
 \wh{\overline{\cM}}{}_{2,L}^{-1} &=  \wh{\overline{\cK}}{}_{2,L}^{-1} +\wh F\,,
 \end{align}
while 
$\wh F$, $\wh G$, $\wh F_3$, $\wh{\overline{\cK}}_{2,L}$, 
and $\wh \cK_{\df,3}$ are defined in Sec.~\ref{sec:formalism}.
The bras and kets on the right-hand side of (\ref{eq:MtoKL}) convert matrices
in the $\{k\ell m i\}$ space into functions of on-shell momenta.
First, there is an implicit combination of the $\ell m$ indices with spherical harmonics,
as in \Cref{eq:Kdfdecomp}.
Then the flavor-matrix structure is removed by combining with
$\bra{\alpha_\cS}=S_{11'} \bra{\alpha}$,
with  $\bra{\alpha}= (1, 1/\sqrt2)$ a row vector in flavor space,
and $S_{11'} = 1 + P_{11'}$, with $P_{11'}$ permuting the momenta of the two
identical particles.
The kets are given by the transposes.

To obtain the infinite-volume amplitude $\cM_3$, 
one sends $L\to\infty$ in \Cref{eq:MtoKL} after first regularizing
the poles in $\wh F$ and $\wh G$ with the $i\epsilon$ prescription. This converts the
matrix structure in the last term in \Cref{eq:MtoKL} into integral equations.  We will not
need the explicit form of these.


The amplitude $\cM_3$ diverges for certain choices of external momenta, including at threshold.
The divergences arise from repeated interactions between alternating pairs
of particles connected by a propagator~\cite{Rubin:1966zz},
and are contained in the first term on the right-hand side of Eq.~(\ref{eq:MtoKL}).
It is therefore useful to introduce a subtracted amplitude, $\cM_{\df,3}$, 
that is free of these divergences.
This was defined for identical particles in Ref.~\cite{\HSQCb},
and for nondegenerate particles in Ref.~\cite{\BSnondegen}.
The corresponding quantity for the $2+1$ theory was not written down explicitly in 
Ref.~\cite{\BStwoplusone}, but its form is immediately clear from Eq.~(\ref{eq:MtoKL}).
It is given by the $L\to\infty$ limit of
\begin{equation}
\cM_{3,\df,L} = \cM_{3,L} - \bra{\alpha_\cS} \wh \cD_L^{(u,u)} \ket{\alpha_\cS}\,,
\label{eq:Mdf3}
\end{equation}
where the second term on the right-hand side removes the divergences in the first.
We stress that the flavor-matrix structure ensures that all choices of two-particle interactions occur.
It then follows that $\cM_{\df,3}$ is given by the $L\to\infty$ limit of
\begin{equation}
\cM_{\df,3,L} =
 \bra{\alpha_\cS} [1/3 - \wh \cD_{23,L}^{(u,u)} \wh F]
\wh \cK_{\df,3} \frac1{1 + \wh F_3 \wh \cK_{\df,3}}
[1/3 - \wh F \wh \cD_{23,L}^{(u,u)}]\ket{\alpha_\cS}\,.
\label{eq:MtoKdfL}
\end{equation}
This gives the integral equations relating $\cM_{\df,3}$ and $\wh\cK_{\df,3}$.

\begin{sloppypar}
We next consider this relation in ChPT. We can calculate $\cM_3$ and thus, using
Eq.~(\ref{eq:Mdf3}), also $\cM_{\df,3}$. If we work at tree level, then, as we show in the main text,
${\cM_{3} \sim \cM_{\df,3} \sim M^2/F^4}$, with $M$ a meson mass. In the standard chiral counting
${\wh \cD_{23,L}^{(u,u)} \sim \cM_2 \sim 1/F^2}$, while $\wh F\sim \wh F_3 \sim \cO(1)$.
It then follows from Eq.~(\ref{eq:MtoKdfL}) that $\Kdf\sim \cM_{\df,3}$.
Using these results, we obtain
\end{sloppypar}
\begin{equation}
\cM_{\df,3}^{\rm LO}
 =
\lim_{L\to\infty}\frac19 \bra{\alpha_\cS} \wh \cK_{\df,3} \ket{\alpha_\cS}\,.
\label{eq:MdftoKdf}
\end{equation}
In addition, at this order, only the first term in the infinite series of subtractions contained
in $\wh \cD_L^{(u,u)}$ appears, so that
\begin{equation}
\cM_{\df,3}^{\rm LO}
 = \cM_3^{\rm LO} + \lim_{L\to \infty} \bra{\alpha_\cS} 
 \wh{\overline{\cM}}{}_{2,L}^{\rm LO} \wh G \wh{\overline{\cM}}{}_{2,L}^{\rm LO} \ket{\alpha_\cS}
 \,.
 \label{eq:Mdf3LO}
 \end{equation}
 We stress that a calculation of $\Kdf$ at NLO in ChPT would involve additional complications.
 Not only would one need to calculate $\cM_3$ to NLO, 
 an arduous task recently completed for the $3\pi^+$ case~\cite{Bijnens:2021hpq},
 one would also have to implement the subtraction at NLO
 and invert the integral operators contained in Eq.~(\ref{eq:MtoKdfL}).
 
All that remains is to explicitly work out the matrix structure in Eqs.~(\ref{eq:MdftoKdf})
and (\ref{eq:Mdf3LO}). For the former this is simple, because
we know from Eq.~(\ref{eq:symmKdf3}) that, after recombination with spherical harmonics,
each entry in $\Kdf$ is given by the same function, up to symmetry factors. 
Furthermore, the symmetries
of $\Kdf$ imply that $S_{11'}$ can be replaced by a factor of $2$. Thus we have that
\begin{equation}
\bra{\alpha_\cS} \to 2 \bra{\alpha}\,,\qquad
\ket{\alpha_\cS} \to 2 \ket{\alpha}\,,\qquad
\wh \cK_{\df,3} \to \ket{\alpha} \Kdf \bra{\alpha}\,.
\end{equation}
Substituting in these results, we obtain
\begin{equation}
\frac19 \bra{\alpha_\cS} \wh \cK_{\df,3} \ket{\alpha_\cS}
= \frac19 2^2 \langle\alpha|\alpha\rangle \Kdf \langle\alpha|\alpha\rangle
= \frac19 2^2 \left(\frac32\right)^2 \Kdf = \Kdf\,.
\end{equation}
Thus, at LO in ChPT we simply have the same result as in the identical-particle
case~\cite{\HHanal},
\begin{equation}
\cM_{\df,3} = \cK_{\df,3}\,.
\end{equation}

Finally, we write out a fully explicit form of $\cM_{\df,3}^{\rm LO}$, Eq.~(\ref{eq:Mdf3LO}).
We will need the result
\begin{align}
\wh{\overline{\cM}}_{2,L} &= {\rm diag}
\left( \overline\cM_{2,L}^{(1)}, \frac12 \overline\cM_{2,L}^{(2)} \right)\,,
\label{eq:M2Lhat}
\\
\left[\overline\cM_{2,L}^{(i)}\right]_{p\ell' m';k\ell m}
&= \delta_{\bm p \bm k} \delta_{\ell' \ell} \delta_{m' m} 2 \omega_p^{(i)} L^3
\cM_{2,\ell}^{(i)}(q_{2,p}^{*(i)})\,,
\end{align}
where $\cM_{2,\ell}^{( i)}$ is the on-shell scattering amplitude in the $\ell$th wave
of the pair that remain when flavor $i$ spectates.
One simplification is that the subtraction term in Eq.~(\ref{eq:Mdf3LO})
involves only $s$-wave interactions, so no sum over $\ell$ is required. 
In principle, one might have expected a $p$-wave contribution to $K\pi$ scattering, 
but this does not appear until NLO in ChPT.
We now take the infinite-volume limit,
following the approach described in Sec.~VII of Ref.~\cite{\BSnondegen}, finding
\begin{equation}
\cM_{\df,3}^{\rm LO}(\{\bm p\}, \{\bm k\}) = \cM_3^{\rm LO}(\{\bm p\}, \{\bm k\}) 
-\cD^{\rm LO}(\{\bm p\}, \{\bm k\}) \,,
\label{eq:Mdf3final}
\end{equation}
where the subtraction term is 
\begin{multline}
\cD^{\rm LO}(\{\bm p\}, \{\bm k\}) =
- S_{11'} \cM_{2,s}^{(1),\rm LO}(q_{2,p_1}^{*(1)}) 
G^{(12),\infty}_s(\bm p_1,\bm k_2) \cM_{2}^{(2,\rm LO)}(q_{2,k_2}^{*(2)})
\\
- \cM_{2,s}^{(2),\rm LO}(q_{2,p_2}^{*(2)}) 
G^{(21),\infty}_s(\bm p_2, \bm k_1) \cM_{2,s}^{(1),\rm LO}(q_{2,k_1}^{*(1)}) S_{11'}
\\
-S_{11'} \cM_{2,s}^{(1),\rm LO}(q_{2,p_1}^{*(1)})
G^{(11),\infty}_s(\bm p_1,\bm k_1) \cM_{2,s}^{(1),\rm LO}(q_{2,k_1}^{*(1)}) S_{11'}
\,.
\label{eq:subLO}
\end{multline}
Here $\{\bm p\}\equiv \{\bm p_1,\bm p_{1'}, \bm p_2\}$ and
$\{\bm k\}\equiv \{\bm k_1,\bm k_{1'}, \bm k_2\}$ are the sets of on-shell
momenta for the final and initial states, respectively, and
\begin{align}
G^{(ij),\infty}_s(\bm p, \bm r) &= \frac{1}{b_{ij}^2 - m_k^2 + i\epsilon}\,,
\end{align}
with $b_{ij}$ defined as in Eq.~(\ref{eq:bij}).
We observe the cancellation of all symmetry factors present in $\wh G$
and $\wh{\overline{\cM}}_{2,L}$
[see Eqs.~(\ref{eq:Ghat}) and (\ref{eq:M2Lhat}), respectively], 
which requires noting that one can replace $S_{11'}$ with $2$ when applied to
 $\overline\cM_2^{(2)}$, as it is already symmetric under $1\leftrightarrow 1'$.
In total there are $4+2+2=8$ subtraction diagrams, four with exchange of the particle
with flavor 2, and four with exchange of the particle with flavor 1.
Finally we note that the LO subtraction does not depend on the cutoff functions, a result that
does not hold at higher orders in ChPT.

\section{Comments on code}
\label{app:code}
%

The goal of this section is to provide simple examples of the 
content and usage of the public code shared along this paper in the repository~\cite{coderepo}. 
Specifically, we discuss the scripts \verb+test.py+ and \verb+solution.py+, 
which can be found in the aforementioned repository.

\subsection{Evaluating the QC3}

Let us start with \verb+test.py+. The goal of this script is provide an example of evaluation of the QC3. We need to start by defining the properties of the particles, and the size of the box:
\begin{python}
M1,M2 = [100.,50.]  # The 3-pt. system masses are [M1,M1,M2], e.g. in MeV
M1,M2 = [1.,M2/M1]  # We always rescale by M1 to make everything dimensionless
M12 = [M1,M2]
parity = -1        # Particle parity (-1 for pseudoscalars)                                                                                                                         
L = 5              # Box size (in units of 1/M1)                                                                                                                                                    
\end{python}
as well as choosing the kinematical configuration at which the QC3 is evaluated:
\begin{python}
nnP = [0,0,0]  # 3-pt. FV spatial momentum (integer-valued)                                                                                                                                     
Ecm = 3.1      # Total 3-pt. energy in CM frame (in units of M1)
E = defns.E_to_Ecm(Ecm,L,nnP,rev=True) # Total 3-pt. energy in moving frame
\end{python}
Note that all quantities are expressed in units of \pythoninline{M1}. 

We will now define the parameters of two-particle interactions:
\begin{python}
waves = 'sp'  # Partial waves used for dimers with flavor-1 spectators                                                                                    
              # (flavor-2 spectators only use s-wave)                                                                                                                                     
a_1s = 0.15   # s-wave scattering length for spectator-flavor-1 channel                                                                                                 
r_1s = 0.0    # s-wave effective range for spectator-flavor-1 channel                                                                                                 
a_1p = 0.2    # p-wave scattering length for spectator-flavor-1 channel                                                                                                                   
a_2s = 0.1    # s-wave scattering length for spectator-flavor-2 channel          
\end{python}
and build the corresponding $q^{2\ell+1} \cot \delta_\ell$ functions:
\begin{python}
f_qcot_1s = lambda q2: qcot_fits.qcot_fit_s(q2,[a_1s,r_1s],ERE=True)
f_qcot_1p = lambda q2: qcot_fits.qcot_fit_p(q2,a_1p)
f_qcot_1sp = [f_qcot_1s, f_qcot_1p]
f_qcot_2s = [lambda q2: qcot_fits.qcot_fit_s(q2,[a_2s],ERE=True)]
\end{python}
Finally, we need to choose the parameters of $\Kdf$. There are four to linear order, as explained in \Cref{eq:thexp}.
\begin{python}
K3iso = [200, 400]     # Isotropic term is K3iso[0] + K3iso[1]*\Delta                                                                                                                              
K3B_par = 400          # Parameter for Kdf3^B1 term                                                                                                                                                    
K3E_par = 300          # Parameter for Kdf3^E1 term          
\end{python}

Before evaluating the quantization condition, we can check which values of the spectator momenta are relevant for the present kinematic configuration.
\begin{python}
nnk_list_1 = defns.list_nnk_nnP(E,L,nnP, Mijk=[M1,M1,M2])
nnk_list_2 = defns.list_nnk_nnP(E,L,nnP, Mijk=[M2,M1,M1])
print('flavor 1 spectators:\n', nnk_list_1)
print('flavor 2 spectators:\n', nnk_list_2)
\end{python}
The output of the previous code will be:
\begin{Verbatim}[fontsize=\small]
flavor 1 spectators:
[(0, 0, 0)]
flavor 2 spectators:
[(0, 0, 0), (0, 0, 1), (0, 1, 0), (1, 0, 0), (0, 0, -1), (0, -1, 0), (-1, 0, 0)]
\end{Verbatim}
This indicates that the matrices involved in the quantization condition will have dimension $(1 \times 4+7)=11$. The factor $4$ in the previous count comes from the presence of $s$ and $p$ waves when the spectator is of type 1.

We now turn to the valuation of the quantization condition, which will be stored in the variable \pythoninline{QC3_mat}:
\begin{python}
F3 = F3_mat.F3mat_2plus1(E,L,nnP, f_qcot_1sp,f_qcot_2s,
                         M12=M12,waves=waves)
K3 = K3main.K3mat_2plus1(E,L,nnP, K3iso,K3B_par,K3E_par,
                         M12=M12,waves=waves)
F3i = LA.inv(F3)
QC3_mat = F3i + K3       # QC3 matrix as defined in the paper                                   
\end{python}
We can check its eigenvalues, organized by magnitude from smallest to largest:
\begin{python}
EV_QC3 = sorted(defns.chop(LA.eigvals(QC3_mat).real,tol=1e-9),key=abs)
print('Eigenvalues of the QC:\n', np.array(EV_QC3))
\end{python}
This will result in
\begin{Verbatim}[fontsize=\small]
Eigenvalues of the QC:
[ 60310.715  60500.739  60500.739  60646.741  60646.741  60646.741
  97412.066 200914.502 200914.502 200914.502 270697.19 ]
\end{Verbatim}

Finally, we can project the matrix \pythoninline{QC3_mat} onto the different irreps of the octahedral group $O_h$, and check the eigenvalues in each irrep:
\begin{python}
print('QC3_mat eigenvalues by irrep ({} total)'.format(len(QC3_mat)))
for I in GT.irrep_list(nnP):
  M_I = proj.irrep_proj_2plus1(QC3_mat,E,L,nnP,I,
                               M12=M12, waves=waves, parity=parity)
  if M_I.shape != (0,0):
    M_I_eigs = sorted(defns.chop(LA.eigvals(M_I).real,tol=1e-9), key=abs)
    print('Irrep = ', I)
    print('Eigenvalues = ', np.array(M_I_eigs))
\end{python}
The output of the code will be:
\begin{Verbatim}[fontsize=\small]
QC3_mat eigenvalues by irrep (11 total)
Irrep =  T1g
Eigenvalues =  [ 60646.741  60646.741  60646.741 200914.502 200914.502 200914.502]
Irrep =  A1u
Eigenvalues =  [ 60310.715  97412.066 270697.19 ]
Irrep =  Eu
Eigenvalues =  [60500.739 60500.739]
\end{Verbatim}
Note that the same eigenvalues appear in the projected and unprojected matrices, and that the overall counting matches.

\subsection{Solving the three-particle quantization condition}

Here, we give an example of how to solve the quantization condition. 
This is provided in the script \verb+solution.py+ provided in the repository. 
We omit the discussion of the first part of the script, 
since it is the same as in \verb+test.py+---it deals with defining
the interactions and properties of the particles.

In order to find solutions, 
it is useful to define a function that returns the smallest eigenvalue in magnitude in some irrep:
\begin{python}
def QC3(Ecm, L, nnP, f_qcot_1sp, f_qcot_2s, M12, waves,
        K3iso, K3B_par, K3E_par, parity, irrep):
  E = defns.E_to_Ecm(Ecm,L,nnP, rev=True)
  F3 = F3_mat.F3mat_2plus1(E,L,nnP, f_qcot_1sp,
                           f_qcot_2s, M12=M12,waves=waves)
  K3 = K3main.K3mat_2plus1(E,L,nnP, K3iso,
                           K3B_par,K3E_par, M12=M12,waves=waves)
  F3i = LA.inv(F3)
  QC3_mat = F3i + K3
  QC3_mat_I = proj.irrep_proj_2plus1(QC3_mat,E,L,nnP,irrep,
                                     M12=M12,waves=waves,parity=parity)
  eigvals = sorted(defns.chop(LA.eigvals(QC3_mat_I).real,tol=1e-9), key=abs)
  return eigvals[0]
\end{python}
A solution of the quantization condition is therefore located at an energy for which \pythoninline{eigvals[0]} is zero. 

We now choose the energy levels for which we want the interacting finite-volume energy.
It is often convenient to know all noninteracting CMF energies below some benchmark $E^*_{\rm max}$, including degeneracies and irrep makeup:
\begin{python}
Ecm_max = 3.8
Ecm_free_list = GT.free_levels_decomp_3pt([M1,M1,M2],L,nnP,Ecm_max=Ecm_max,
                                       		  sym='2+1',parity=parity)
for n in range(len(Ecm_free_list)):
  Ecm_free, degen, decomp = Ecm_free_list[n]
  # decomp = list of tuples of the form (irrep, # of copies)
  print('Ecm_free: {:.6f}, degen: {}, irreps: {}'.format(
                                                  Ecm_free, degen, decomp))
\end{python}
The output of the code is
\begin{Verbatim}[fontsize=\small]
Ecm_free: 2.500000, degen: 1, irreps: [('A1u', 1)]
Ecm_free: 3.711938, degen: 3, irreps: [('A1u', 1), ('Eu', 1)]
\end{Verbatim}
We now turn to solving the quantization condition. We start by defining a wrapper that depends only on the CMF energy of the three-particle system. All other variables in the function have been previously defined.
\begin{python}
func = lambda Ecm_arr: QC3(Ecm_arr[0], L, nnP, f_qcot_1sp, f_qcot_2s, M12,
                            waves, K3iso, K3B_par, K3E_par, parity, irrep)
\end{python}
Finally, we can find solutions of \pythoninline{func} using the solver \pythoninline{scipy.optimize.fsolve}. For this, we need to provide an initial guess for the solution. Since the interactions are repulsive, we choose the value of the noninteracting energy plus some positive shift: \pythoninline{Ecm_free+0.06}.
\begin{python}                
for n in range(len(Ecm_free_list)):
  Ecm_free, degen, decomp = Ecm_free_list[n]
  for (irrep, N_copies) in decomp:
    Ecm_test = Ecm_free + 0.06          # Try an energy slightly above Ecm_free
    Ecm_sol = fsolve(func, Ecm_test)[0] # Will only find one solution
    					# (more work needed if N_copies > 1)
    print('irrep: {}, solution: {:.6f}, shift: {:.6f}'.format(irrep, Ecm_sol, 
    							     Ecm_sol-Ecm_free))
\end{python}
For this choice of parameters, the solutions of the quantization condition are:
\begin{Verbatim}[fontsize=\small]
irrep: A1u , solution:  2.556465 , shift:  0.056465
irrep: A1u , solution:  3.767897 , shift:  0.055959
irrep: Eu  , solution:  3.765428 , shift:  0.053489
\end{Verbatim}

\subsection{Further comments}

We also include three scripts showing additional features. Two of them, \verb+solution_ND.py+ and \verb+solution_ID.py+, provide examples for finding solutions of the QC3 for three nondegenerate scalars and three identical scalars, respectively. These scripts are very similar to the script described in the previous subsection. However, only $s$ waves are included in both cases.

The main difference involves the construction of the $F_3$ and $\Kdf$ matrices, which require the usage of appropriately modified functions:
\begin{python}
F3_mat.F3mat_ND(E,L,nnP, f_qcot_1s,f_qcot_2s,f_qcot_3s, M123=[M1,M2,M3])
F3_mat.F3mat_ID(E,L,nnP, f_qcot_s)
K3main.K3mat_ND_iso(E,L,nnP, K3iso, M123=[M1,M2,M3)
K3main.K3mat_ID_iso(E,L,nnP, K3iso)
\end{python}
Note that in the first function one needs three different phase shifts, one for each pairwise interaction, while for the second only one phase shift is needed. 

The last script, \verb+solution_2pt.py+, provides an implementation of the two-particle quantization condition. Only $s$ waves are included for the case of identical particles, and $s$ and $p$ waves for nonidentical scalars.

\bibliographystyle{JHEP}      
\bibliography{ref.bib}

\end{document}